\documentclass{aa}  

\usepackage{graphicx}
\usepackage{txfonts}
\usepackage[colorlinks=true, allcolors=blue]{hyperref}
\begin{document}

\title{NGC~3259: A Signal for an Untapped Population of Slowly Accreting Intermediate-Mass Black Holes}

\author{Kirill A. Grishin\inst{1, 2}
        \and
        Igor V. Chilingarian\inst{3, 2}
        \and
        Francoise Combes\inst{4}
        \and
        Franz E. Bauer\inst{5,6}
        \and
        Victoria A. Toptun\inst{7}
        \and
        Ivan Yu. Katkov\inst{8,9,2}
        \and 
        Daniel Fabricant\inst{3}
        }

 \institute{Universite Paris Cite, CNRS, Astroparticule et Cosmologie, F-75013 Paris, France
            \and
            Sternberg Astronomical Institute, M.V. Lomonosov Moscow State University, 13 Universitetsky prospect, 119992 Moscow, Russia
            \and
            Center for Astrophysics | Harvard \& Smithsonian, 60 Garden St., Cambridge, MA 02138, USA
            \and
            Observatoire de Paris, LUX, Collège de France, CNRS, PSL University, Sorbonne University, 75014, Paris
            \and
            Instituto de Astrofísica and Centro de Astroingeniería, Facultad de Física, Pontificia Universidad Católica de Chile, Campus San Joaquín, Av. Vicuña Mackenna 4860, Macul Santiago, Chile
            \and
            Millennium Institute of Astrophysics, Monseñor Nuncio Sótero Sanz 100, Of 104, Providencia, Santiago, Chile
            \and
            European Southern Observatory, Karl Schwarzschildstrasse 2, 85748, Garching bei München, Germany
            \and
            New York University Abu Dhabi, P.O. Box 129188, Abu Dhabi, UAE
            \and
            Center for Astrophysics and Space Science (CASS), New York University Abu Dhabi, P.O. Box 129188, Abu Dhabi, UAE            
            }

 
\abstract{Low-mass active galactic nuclei (AGNs) can provide important constraints on the formation and evolution of supermassive black holes (SMBHs), a central challenge in modern cosmology. To date only small samples of intermediate-mass black holes (IMBHs, $M_{BH} < 10^5 M_{\odot}$) and `lesser' supermassive black holes (LSMBH, $M_{BH} < 10^6 M_{\odot}$) have been identified. Our present study of NGC~3259 at $D=27$~Mpc with the Binospec integral field unit (IFU) spectrograph complemented with Keck Echelle Spectrograph and Imager observations demonstrates the need for and the power of the spectroscopic follow-up. NGC~3259 hosts a black hole with a mass of $M_{BH} = (1.7 - 4.1) \times 10^5 M_{\odot}$, inferred from multi-epoch spectroscopic data, that accretes at 1\% of the Eddington limit as suggested by the analysis of archival XMM-Newton observations. It is the second nearest  low-mass AGN after the archetypal galaxy NGC~4395. The spectroscopic data reveals a variable broad $H\alpha$ profile that is likely the result of asymmetrically distributed broad-line region (BLR) clouds or BLR outflow events.  X-ray observations and the absence of an optical power-law continuum suggest partial obscuration of the accretion disk and hot corona by a dust torus.
We estimate that the Sloan Digital Sky Survey (SDSS) could only detect similar objects to $D=35$~Mpc.  
A detailed photometric analysis of NGC~3259 using HST images provides a central spheroid stellar mass estimate 25 times lower than expected from the $M_{BH} - M^*_{sph}$ relation, making this galaxy a strong outlier. This discrepancy suggests divergent growth pathways for the central black hole and spheroid, potentially influenced by the presence of a bar in the galaxy. Finally, we demonstrate that the DESI and 4MOST surveys will detect low-accretion rate IMBHs and LSMBHs and the sensitivity of future X-ray instruments (such as AXIS and Athena) will secure their classification.}



\date{Received ; accepted }

   \maketitle
%

\section{Introduction}

Supermassive black holes (SMBHs) are thought to grow through two main channels: gas accretion~\citep{2012Sci...337..544V} and mergers~\citep{2005LRR.....8....8M}. The latter channel establishes a scaling relation between the galaxy bulge mass (central spheroid) and the central black hole mass ($M^*_{sph}$-$M_{BH}$), that reflects the coevolution of these two components~\citep{2013ARA&A..51..511K}. SMBHs and massive galaxies obey this scaling relation, indicating that galaxy mergers are the dominant growth mechanism for SMBHs in this mass range~\citep{2015ApJ...798...54G, 2023MNRAS.518.2177G, 2024MNRAS.535..299G}.  At lower masses, gas accretion could play a more significant role in black hole (BH) growth, changing the coevolution of the BH and the central spheroid. However, studies of lower-mass supermassive black holes (lesser SMBH; LSMBHs, $2\times10^5 < M_{BH} < 10^6 M_{\odot}$) and intermediate-mass black holes (IMBHs, $100 <M_{BH} < 2\times 10^5 M_{\odot}$) suggest that they follow the SMBH $M^*_{sph}$-$M_{BH}$ relation~\citep{2007ApJ...670...92G, 2018ApJ...863....1C}.

The selection biases for the current LSMBH and IMBH samples are poorly understood.  Most of the known \textit{bona fide} IMBHs with X-ray counterparts have high accretion rates and significant luminosities in broad H$\alpha$ and X-ray emission, enhancing their detectability. A population of low-accretion-rate LSMBHs may be missing from current samples.  Observations of higher-mass SMBHs indicate that the volume density of low Eddington ratio ($\lambda$) of $10^{-2}$ SMBHs is 100--10,000 times higher than the volume density of $\lambda=1$~\citep{2017ApJ...845..134W, 2022ApJS..261....9A} SMBHs.  If LSMBHs and IMBHs follow an Eddington ratio distribution similar to SMBHs, a substantial population of these lower-mass black holes likely remains undetected because of observational limitations.

Central black hole growth may proceed by different processes in low-mass AGNs with low accretion rates than in SMBHs, affecting both the growth of the central BH and the properties of a central spheroid. For example, in the absence of major mergers, bars that funnel gas toward the galaxy center may predominantly drive black hole growth through gas accretion. This inflowing gas can fuel star formation in a pseudo-bulge~\citep{2015MNRAS.454.3641F}. After the gas within the inner ring is fully depleted, low accretion rates may prevent the detection of these low-mass AGNs.  The pseudo-bulge may not have undergone the same co-evolutionary processes as the central black hole and may not follow the $M^*_{sph}$-$M_{BH}$ relation observed for SMBHs.

NGC~3259, an AGN powered by a slowly accreting LSMBH with $M_{BH} = (1.7 - 4.1) \times 10^5 M_{\odot}$, provides an opportunity to explore the detectability of LSMBHs and IMBHs with low accretion rates. The low-mass end of the black hole mass function constrains formation scenarios of high-redshift SMBHs. Theory proposes two primary pathways for SMBH seed growth: (i) through collisions and core collapse in star clusters containing Population III stars~\citep{2004Natur.428..724P} and (ii) rapid infall and direct collapse of gas clouds~\citep{1994ApJ...432...52L, 2006MNRAS.370..289B}. The second mechanism predicts a gap in the present-day black hole mass function, allowing us to differentiate between the two formation scenarios.

Current and planned spectroscopic surveys: the Dark Energy Spectroscopic Instrument \citep[DESI;][]{2016arXiv161100036D} and the 4-metre Multi-Object Spectroscopic Telescope \citep[4MOST;][]{2019Msngr.175....3D}, will expand spectroscopic samples of galaxies in number, redshift, and spectral resolution compared to the existing SDSS \citep[][]{2000AJ....120.1579Y} and LAMOST \citep[][]{2012RAA....12..723Z} datasets.   The newer surveys will enable for more systematic studies of AGNs powered by LSMBHs in a cosmological context by providing more complete information about the broad emission line component. These studies will validate SMBH growth channels in the late stages of the Universe and constrain black hole seed formation in the early Universe~\citep{2008ApJ...676...33D, 2023MNRAS.525..969S, 2025MNRAS.536..851C}. A detailed study of selection biases will inform the new surveys.

We adopt $\Lambda CDM$ cosmology with $\Omega_M = 0.3$, $\Omega_{\Lambda} = 0.7$, $h = 0.72$, and $\sigma_8 = 0.8$. All magnitudes are presented in the AB system~\citep{1983ApJ...266..713O}, and all uncertainties are quoted at the 1$\sigma$ level unless stated otherwise.  

The Sloan Digital Sky Survey spectrum\footnote{\url{https://rcsed2.voxastro.org/data/galaxy/20295}} for NGC~3259 gives a redshift of $z=0.0056$ ($v=1678$~km~s$^{-1}$, or $D = 23.3$~Mpc). The redshift-based distance estimate may be inaccurate due to peculiar velocities of up to a few hundreds km~s$^{-1}$ because NGC~3259 is in a galaxy group.  The Tully--Fisher relation in the Spitzer bands~\citep{2016AJ....152...50T} yields a distance modulus of $(m-M) = 32.63 \pm 0.30$~mag or $D = 33.5 \pm 4.3$~Mpc. The NAM model\footnote{\url{https://edd.ifa.hawaii.edu/NAMcalculator/}}~\citep{2017ApJ...850..207S}, as implemented in the Cosmicflows-3 distance calculator~\citep{2020AJ....159...67K}, gives a distance to NGC~3259 of 27.0~Mpc that we adopt for this paper. This makes NGC~3259 a host of the second most nearby AGN powered by a low-mass black hole after the archetype galaxy NGC~4395 \citep{1989ApJ...342L..11F,1997ApJS..112..315H,1997ApJS..112..391H} located 4.8~Mpc away.


\begin{figure}[]
    \centering
    \includegraphics[width=\hsize]{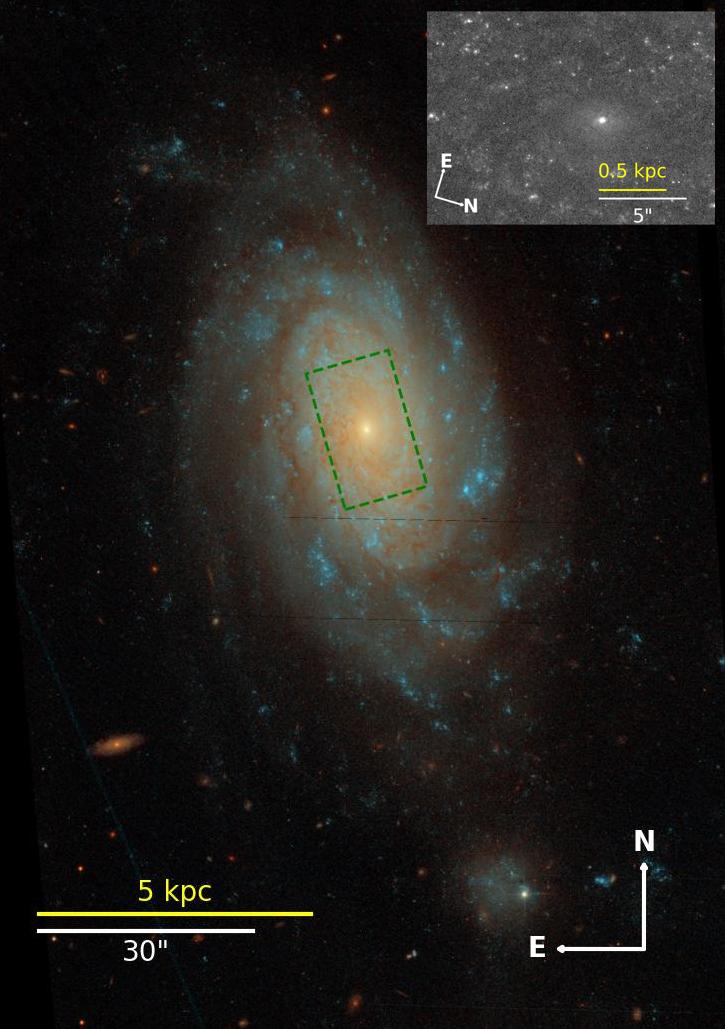}
    \caption{Colour image of NGC~3259 produced by combining \textit{F435W} and \textit{F814W} {\it HST} ACS/WFC images. The green dashed rectangle shows the Binospec IFU field of view and the inset is the portion of the \textit{F330W} \textit{HST ACS/HRC} image in the Binospec IFU field of view. }\label{fig:acs_rgb}
\end{figure}

\section{New and archival observations}

Our study of NGC~3259 uses archival and recently obtained spectroscopic and imaging datasets. In the following subsections, we describe these observations.

\subsection{Hubble Space Telescope spectroscopy}
NGC~3259 was observed with the Space Telescope Imaging Spectrograph~\citep[STIS,][]{1998PASP..110.1183W} on 2000/Jul/09 as part of a snapshot program to assess the $M^*_{sph}$-$M_{BH}$ relation for spiral galaxies (HST-SNAP-8228, PI: D. Axon). The observations used a 52$\times$0.2 arcsec slit with the \textit{G750M} grating, providing a wavelength coverage from 6480 to 7058~\AA~and a spectral resolution of $R\sim10000$ for a point source. Of the six frames obtained for this target, only two showed a clear trace of the target spectrum and were used for spectral extraction. The total exposure time for the two frames is 860~s.  The two fully reduced and calibrated frames, available in the MAST archive, were coadded using a custom cosmic-ray rejection procedure that rejects pixels with flux exceeding that of the second frame by more than $10^{-15}~\mathrm{erg\, cm^{-2} s^{-1} arcsec^{-2}}$.

NGC~3259 was later observed in the far-ultraviolet (FUV) on 2011/Oct/24 with the Cosmic Origins Spectrograph~\citep[COS,][]{2012ApJ...744...60G} as part of the program ``Low-Mass Black Holes and CIV in Low-Luminosity AGN'' (HST-GO-12557, PI: K.~Gultekin).  The 2324~s observation used a circular $d=2.5$~arcsec aperture and the \textit{G160M} grating with a 1623~\AA\ central wavelength.  This configuration provides wavelength coverage from 1432 to 1798~\AA\ with a gap between 1606 and 1625~\AA\ at a spectral resolving power $R=13000-20000$. We downloaded an extracted, fully reduced flux calibrated one-dimensional spectrum from the MAST archive.

\subsection{Integral field unit spectroscopy with Binospec/MMT}
\label{sec:bino_obs}
We observed NGC~3259 using the Binospec spectrograph~\citep{2019PASP..131g5004F} at the 6.5-m MMT in IFU mode~\citep{2025PASP..137a5002F}. The nucleus of NGC~3259 was centered in the 16$\times$12~arcsec$^2$ field of view sampled by 0.68~arcsec diameter hexagonal spaxels\footnote{Here we refer to the circumcircle with a radius equal to the hexagon side length.} aligned with the major axis of the galaxy at a position angle (PA) of 16 degrees (Fig.~\ref{fig:acs_rgb}). The Binospec IFU includes dedicated sky fibers for accurate background subtraction 5~arcmin away from the source. To perform the absolute flux calibration and correct for atmospheric absorption, we also observed the A0V-type star \object{HIP~56147} ($V=7.39$~mag) on the same night at a similar airmass.

NGC~3259 was observed on January 15th, 2024 during the science verification observing run of Binospec-IFU, with a total integration time of 45~min split into individual 15~min long exposures. We used a 600~g mm$^{-1}$ grating, which provided a spectral resolving power of $R \sim $ 4000--5000 in the wavelength range 4710--7170~\AA. These data were collected under good transparency conditions during gray time, with seeing of 0.8--1.1~arcsec. Before the observations we obtained arc lamp and flat field exposures for wavelength calibration and sensitivity correction.

The spectra were reduced using a version of the Binospec pipeline \citep{2019PASP..131g5005K} updated for the reduction of IFU data~\citep{2025PASP..137a5002F}. The pipeline produces a flux-calibrated, sky-subtracted, rectified, and wavelength-calibrated 3D datacube with associated flux and error values. The wavelength axis is linearly sampled with a step size of 0.6~\AA~pix$^{-1}$. We used extracted spectra of an A0V star to construct a telluric absorption model.

For absolute flux calibration we divided a resampled and smoothed Binospec spectrum of \object{HIP~56147} by the low resolution GAIA DR3 BP/RP spectrum of the same star~\citep{2023A&A...674A...1G} to derive a polynomial sensitivity curve.

\subsection{Keck ESI spectroscopy}
NGC~3259 was observed with the Echellette Spectrograph and Imager~\citep[ESI,][]{2002PASP..114..851S} in 2003--2004~\citep{2008AJ....136.1179B}, but this dataset is not available in the Keck Observatory Archive (KOA). Therefore, we re-observed NGC~3259 on 27 March 2023 under dark-sky conditions with intermittent thin cirrus clouds and median seeing of 1~arcsec (NOIRLab program ID 2023A-914712; Keck program ID 2023A-R100). We used the echellette spectroscopic mode with a 0.75-arcsec-wide slit, providing wavelength coverage from 3927 to 11068~\AA~and a mean spectral resolution of R$\approx$8200. We obtained three exposures for a total integration time of 1800~sec. For sky background subtraction, we used two 300-sec exposures 120~arcsec West of the galaxy center, one before and one after the sequence of science exposures. For absolute flux calibration and telluric absorption correction, we observed the A0V star HD~71906 ($V=6.17$~mag) at the same airmass as the science target 30~min prior to the science observations. We performed wavelength calibration using spectra of Hg, Xe, and Cu arc lamps taken during the day.

Spectra were reduced using our ESI pipeline\footnote{\url{https://bitbucket.org/chil_sai/mage-pipeline/src/esi/}} \citep{2020ASPC..522..623C}. The pipeline generates flux-calibrated, sky-subtracted, rectified, wavelength-calibrated, telluric-corrected, and coadded 2D frames with flux and flux uncertainty maps. The weights of the two sky frames were manually adjusted to optimize the residuals of the night sky emission lines. The reduced 2D spectrum has a linearly spaced wavelength grid with a step size of 0.2~\AA~pix$^{-1}$.
We corrected for light loss at the slit using the spectrum of HD~71906.

Our absolute flux calibration procedure is similar to that for Binospec data (Section~\ref{sec:bino_obs}). However, due to the absence of a GAIA calibration star spectrum, we used an A0V synthetic template spectrum from the {\sc Phoenix} library \citep{2013A&A...553A...6H}.  We normalized the synthetic spectrum with HD~71906 near-infrared magnitudes.  The estimated absolute flux calibration accuracy is $\sim$15\%\ due to seeing variations and passing cirrus clouds.

\subsection{Hubble Space Telescope images}
NGC~3259 was observed on June 09, 2002 with the ACS camera onboard the {\it Hubble} Space Telescope ({\it HST}) as part of a galaxy bulge formation program (Proposal ID: 9395, PI: M. Carollo). Images were obtained in three filters: \textit{F330W}, \textit{F435W}, and \textit{F814W}, with the \textit{F330W} dataset taken in ACS/HRC mode (Fig.~\ref{fig:acs_rgb}). The total exposure times for the \textit{F330W} and \textit{F814W} filters are 2760 seconds and 1200 seconds, respectively.  We used publicly available datasets provided by the Hubble Legacy Archive (HLA)\footnote{\url{https://hla.stsci.edu}}.

\subsection{XMM-Newton X-ray observations}

\begin{figure*}
    \centering
    \includegraphics[width=0.46\hsize]{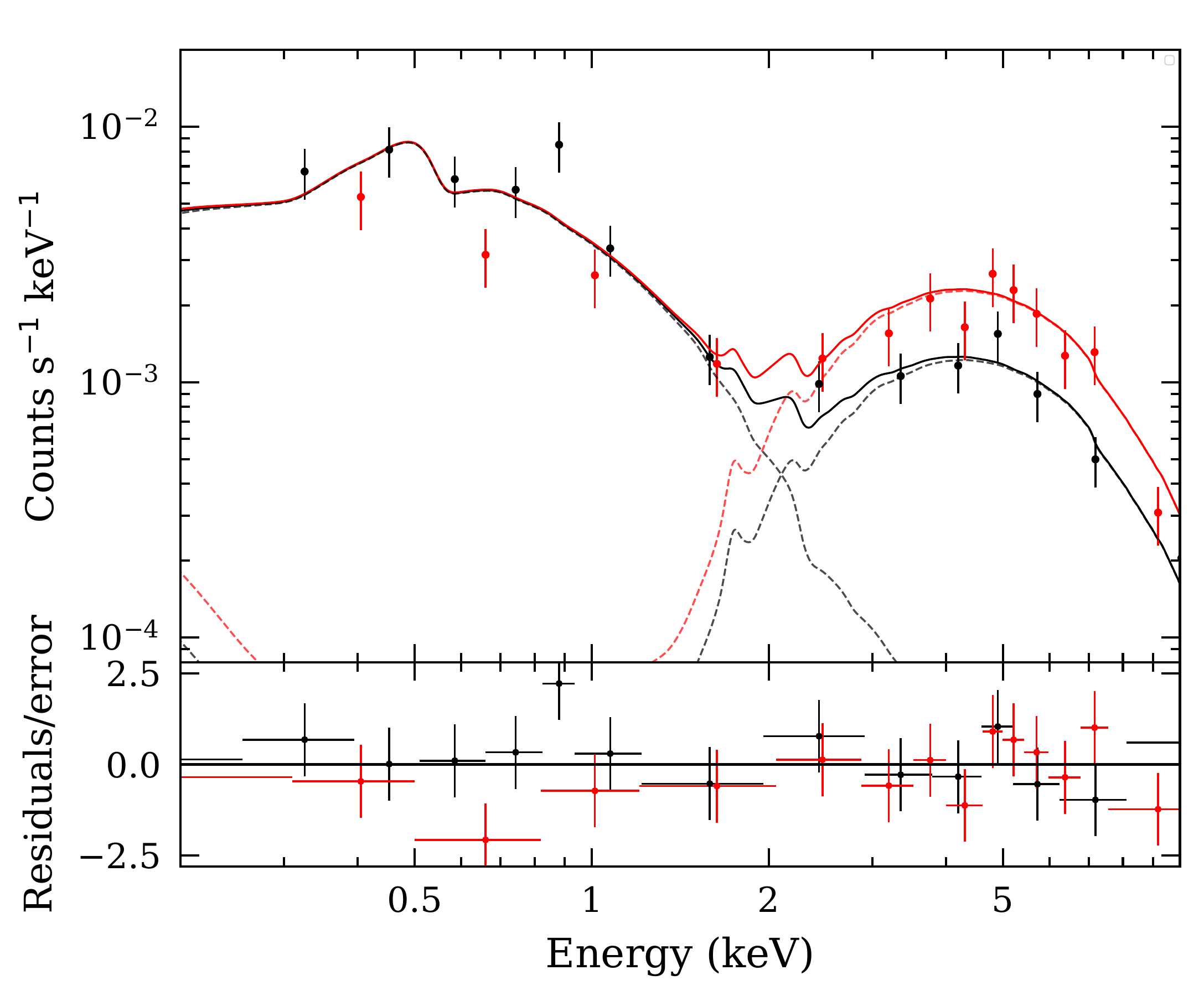}
    \includegraphics[width=0.53\hsize]{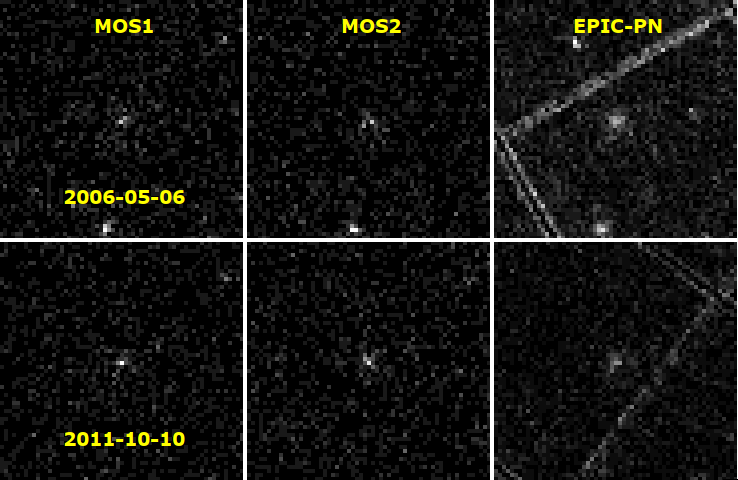}
    \caption{NGC~3259 X-ray data. Left: X-ray spectral models for the {\it XMM-Newton} datasets, obtained in 2006 (black) and 2011 (red). The residuals, divided by the error, are shown in the bottom panel. Right: X-ray images of NGC~3259 from the  {\it XMM-Newton} MOS1,2 and PN instruments.} \label{fig:xray_spec}
\end{figure*}

NGC~3259 was observed twice with {\it XMM-Newton}: once in 2006 (PI: A. Barth; exposure time: 24 ksec., Date: 2006-05-06, Target name: 103234.85+650227.9, Obs. ID: 0400570401) and again in 2011 (PI: E. Cackett; exposure time: 22 ksec, Date: 2011-10-10, Target name: SDSSJ103234, Obs. ID: 0674810701), using the MOS and PN instruments.  We retrieved the two datasets from the XMM-Newton Science Archive (XSA) and performed data reduction following the procedures outlined in XMM SAS~\citep[][]{2004ASPC..314..759G}. The key steps included: (i) Generation of calibrated and concatenated event lists using the {\sc emproc} and {\sc epproc} utilities;\footnote{\url{https://www.cosmos.esa.int/web/xmm-newton/sas-thread-epic-reprocessing}} (ii) Event list cleaning to exclude periods of high background by applying a count rate filtering threshold\footnote{\url{https://www.cosmos.esa.int/web/xmm-newton/sas-thread-epic-filterbackground}}. In the calibrated event lists, NGC~3259 shows a strong X-ray detection at the position of RA=10:32:34.78 and Dec=+65:02:26.7. These coordinates match the optical nuclear point source position in the {\it HST} images with a position error of 0.48 arcsec, within the $3\sigma$ uncertainties of the XMM detection, (Fig. \ref{fig:xray_spec}). After identifying the X-ray counterpart, we extracted spectra using a circular aperture with a radius of 13.4 arcsec. Background spectra were extracted from circular regions in each dataset, chosen to avoid detector gaps and bad pixel columns. We combined the extracted spectra from the EPIC-pn, EPIC-MOS1, and EPIC-MOS2 detectors, using the {\sc epicspeccombine} routine to achieve a higher signal-to-noise ratio for each exposure.

\section{Data analysis and results}

\subsection{Full spectral fitting with emission lines}
\label{sec:spectral_fit}
We applied full spectral fitting to the spectroscopic datasets for NGC~3259 using the {\sc NBursts} IDL package~\citep{2007MNRAS.376.1033C,2007IAUS..241..175C}. This package simultaneously models the stellar population continuum and emission lines from line-of-sight velocity distributions, LOSVDs. The ESI 1-D spectrum was modeled using X-Shooter simple stellar population (SSP) models~\citep{2022A&A...661A..50V} with a spectral resolution of $R \sim 10000$.  This procedure allows estimates of stellar velocity dispersions down to 20~km~s$^{-1}$~\citep{2020PASP..132f4503C}. Our model includes a 15-th degree polynomial multiplicative continuum.  The additive continuum from an AGN was modeled with a second-degree polynomial.

\begin{figure*}[h]
    \centering
    \includegraphics[width=0.245\textwidth,trim={1cm 1cm 0 0},clip]{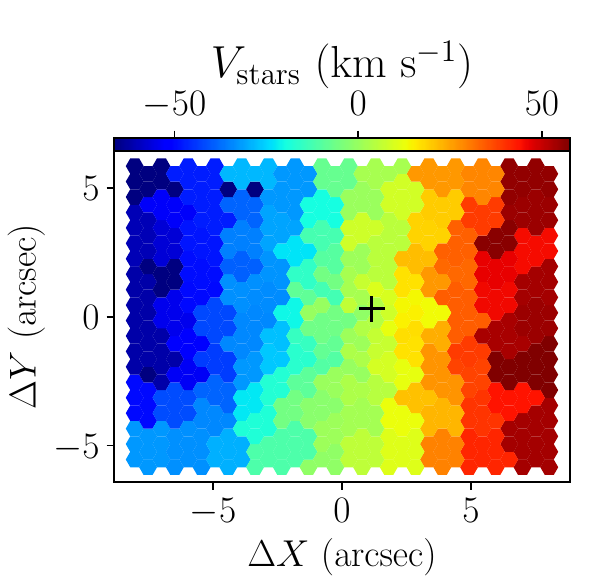}
    \includegraphics[width=0.245\textwidth,trim={1cm 1cm 0 0},clip]{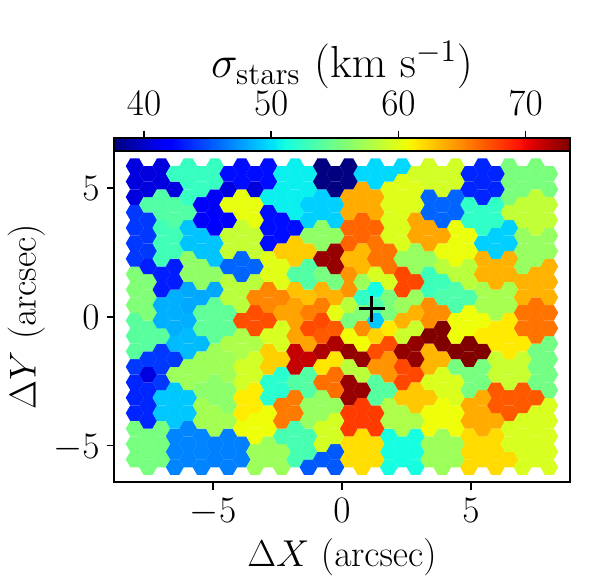}
    \includegraphics[width=0.245\textwidth,trim={1cm 1cm 0 0},clip]{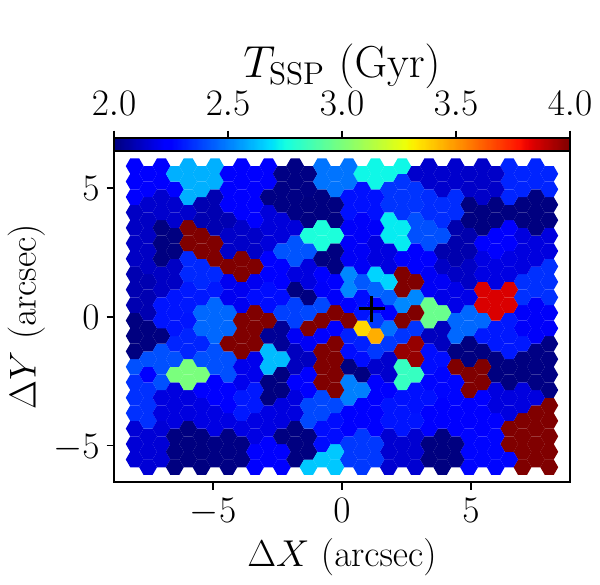}
    \includegraphics[width=0.245\textwidth,trim={1cm 1cm 0 0},clip]{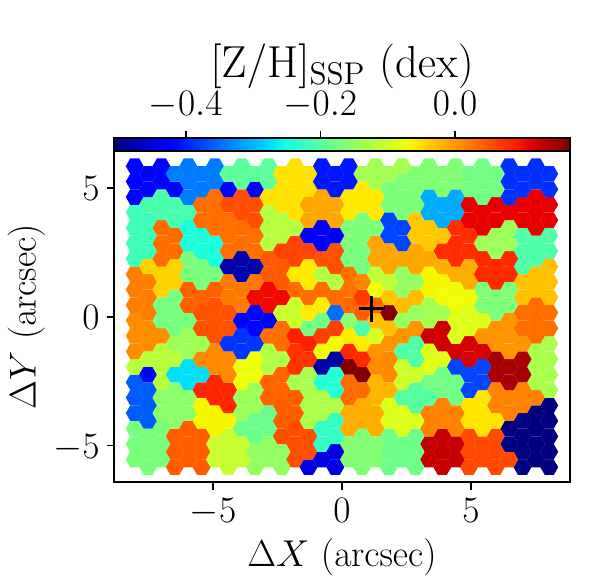}\\
    \includegraphics[width=0.245\textwidth,trim={1cm 1cm 0 0},clip]{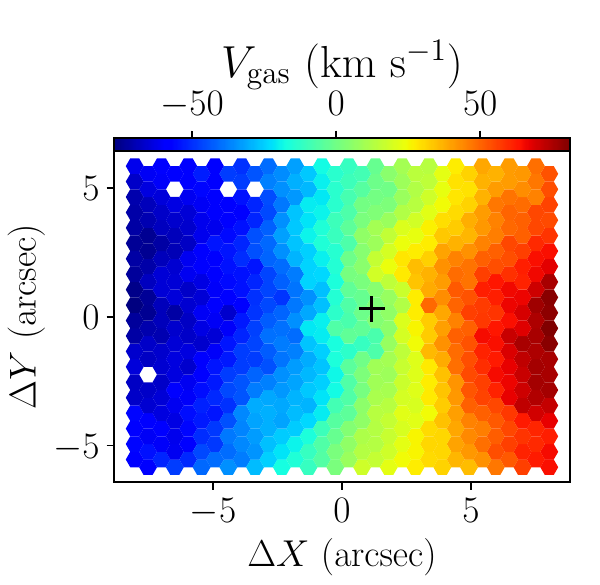}
    \includegraphics[width=0.245\textwidth,trim={1cm 1cm 0 0},clip]{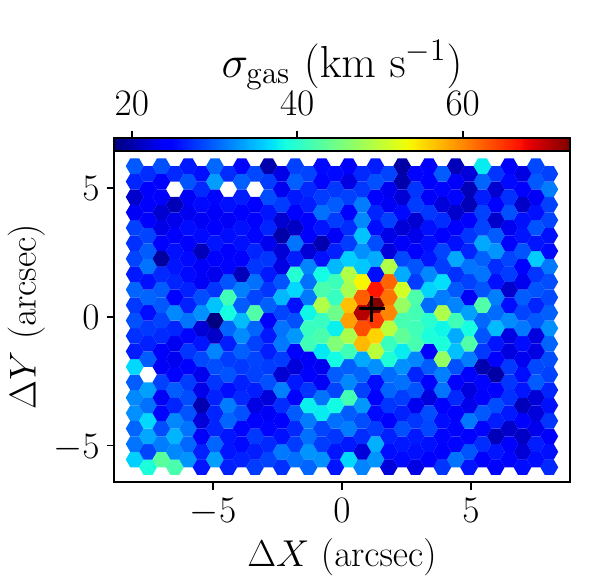}
    \includegraphics[width=0.245\textwidth,trim={1cm 1cm 0 0},clip]{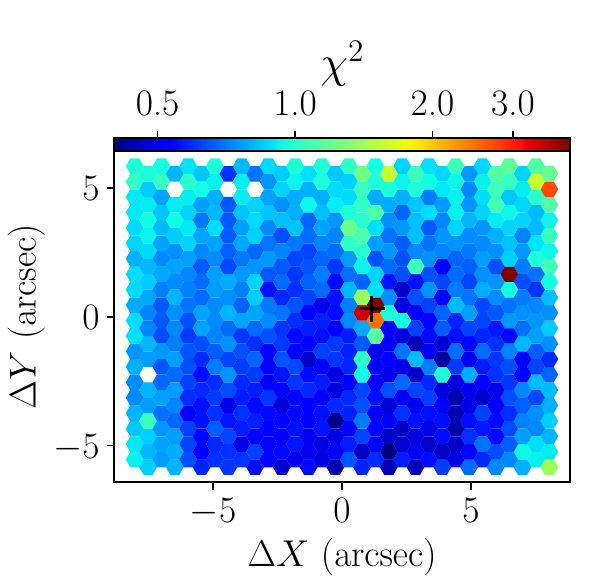}
    \includegraphics[width=0.245\textwidth,trim={1cm 1cm 0 0},clip]{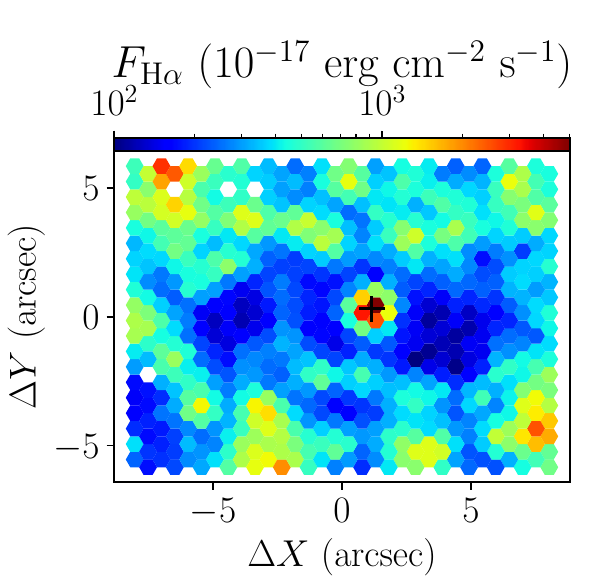}\\
    \includegraphics[width=0.245\textwidth,trim={1cm 1cm 0 0},clip]{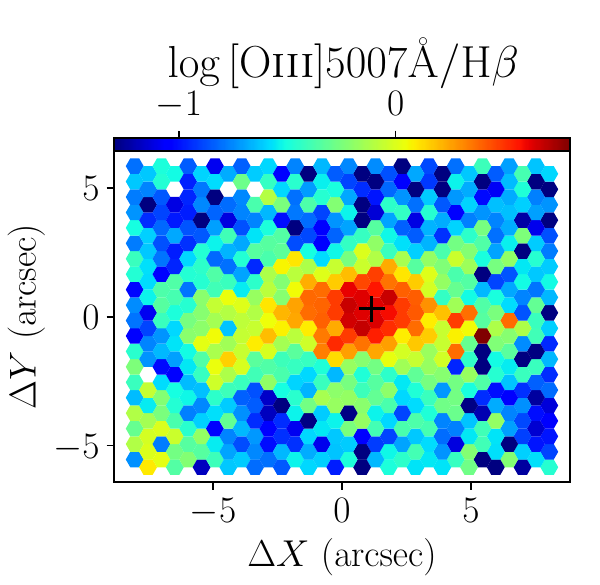}
    \includegraphics[width=0.245\textwidth,trim={1cm 1cm 0 0},clip]{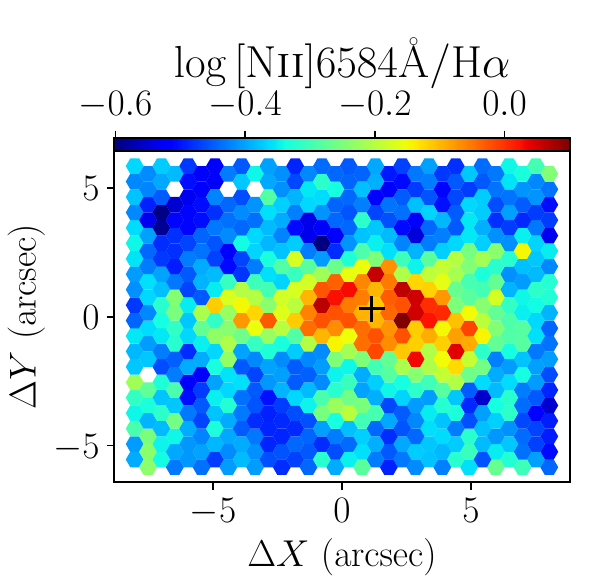}
    \includegraphics[width=0.245\textwidth,trim={1cm 1cm 0 0},clip]{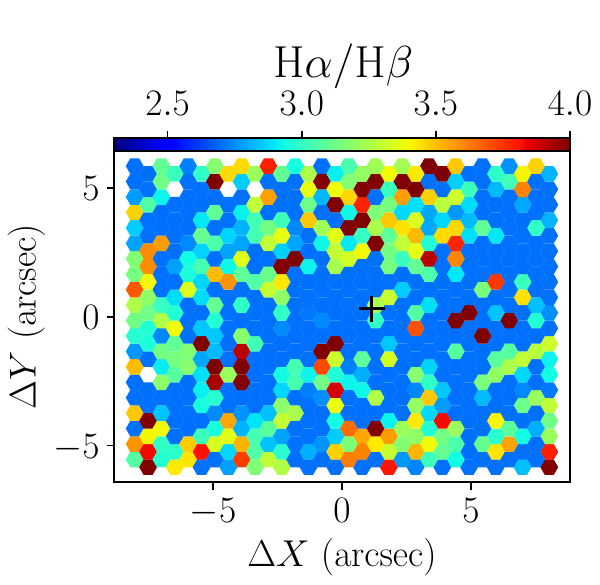}
    \includegraphics[width=0.245\textwidth,trim={1cm 1cm 0 0},clip]{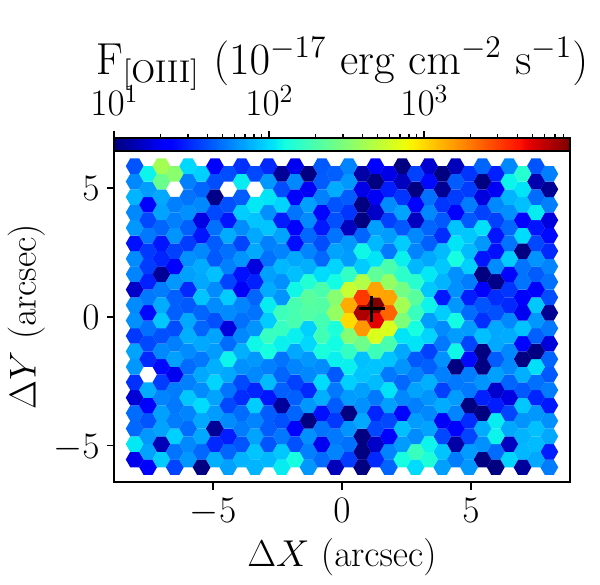}\\
    \includegraphics[width=0.215\textwidth]{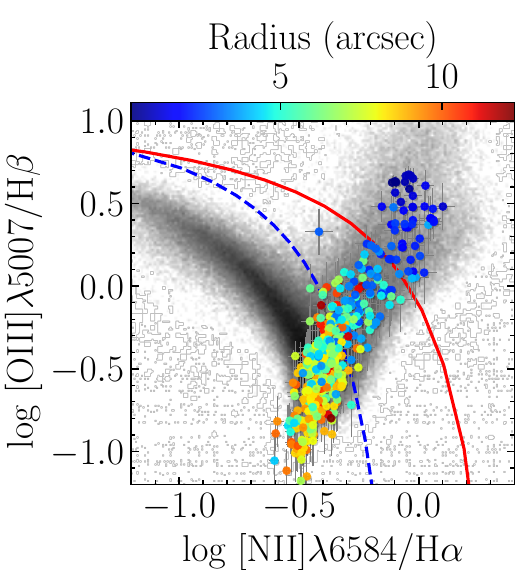}
    \includegraphics[width=0.215\textwidth]{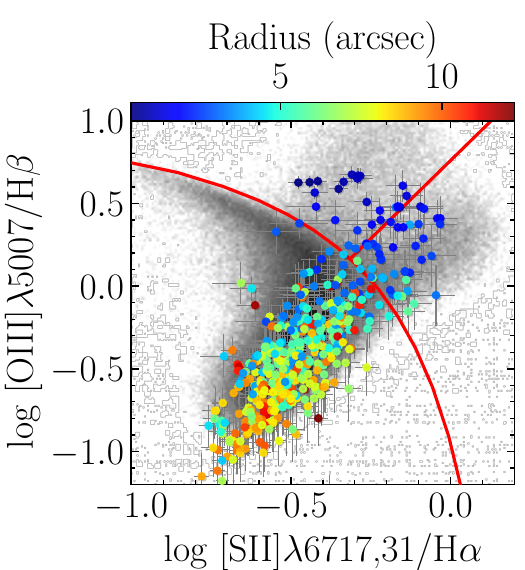}
    \includegraphics[width=0.275\textwidth]{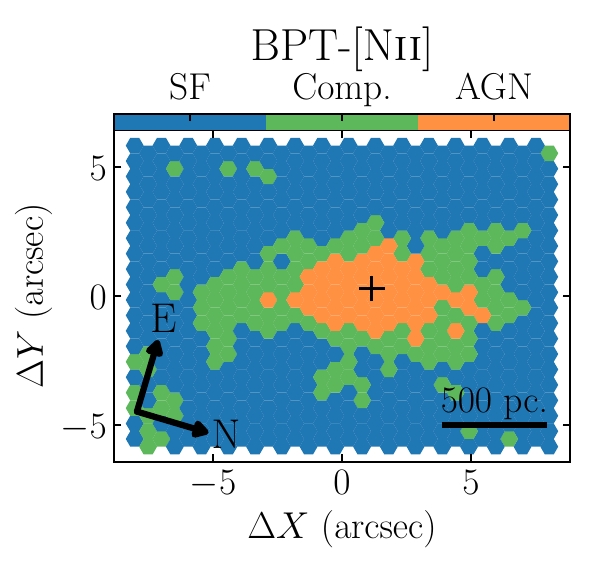}
    \includegraphics[width=0.275\textwidth]{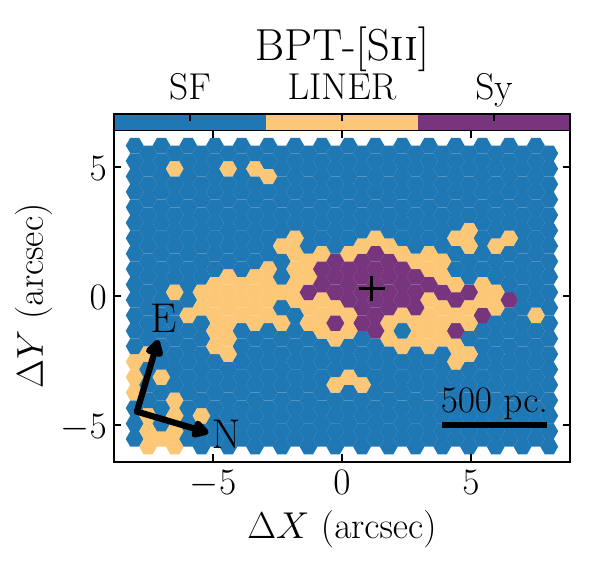}
    \caption{Analysis results from the NGC~3259 Binospec IFU datacube. First (top) row: 1. Stellar population velocity ($V_{stars}$); 2. Stellar population velocity dispersion ($\sigma_{stars}$); 3. Simple stellar population age ($T_{SSP}$); 4. SSP ($[Z/H]_{SSP}$) metallicity. Second row: 1. $\rm{H\alpha}$ emission line velocity ($V_{gas}$); 2. $\rm{H\alpha}$ emission line velocity dispersion ($\sigma_{gas}$); 3. Reduced $\chi^2$ in each spaxel; 4. $\rm{H\alpha}$ emission line flux. Third row: emission line ratios: 1. $\rm{[O\textsc{iii}] /H\beta}$; 2. $\rm{[N\textsc{ii}] / H\alpha}$; 3. Balmer decrement for the narrow line component, $H\alpha/H\beta$ and 4. $\rm{[O\textsc{iii}]}$ emission line flux . Fourth (bottom) row: BPT diagrams (1 and 2) with position of each spaxel with color-coded deprojected distance from the galaxy center; 3-4: color-coded classification of spaxels according to their position on the BPT diagram. On each panel in rows 1-3 we show the center of the broad-line point source with a black cross.
    } \label{fig:bino_maps}
\end{figure*}

For the Binospec datacube, we optimally extracted a 1-D spectrum using a 2-D Gaussian profile in the spectral slices for the [O{\sc iii}] emission line as extraction weights~\citep{1986PASP...98..609H}. This 1-D spectrum was then modeled similarly, with second-degree additive and a 15th-degree multiplicative continua.  To obtain reliable values of the broad-line component parameters it is necessary to extract reliable measurements of broad emission line components from the datacube. Extracting a spectrum within an aperture or using optimal extraction increases the contribution from the host galaxy, while a spectrum from a single central spaxel does not contain the full broad-line component flux.  We analyze the emission spectra from 42 spatial elements within a radius of 2~arcsec from the galaxy center.  We represent the broad component as a 3D Gaussian with two spatial coordinates (essentially a point spread function) and a spectral coordinate. In Appendix~\ref{app:green} we provide the details about precise modeling of the 2D Gaussian for the IFU data. Each element contains a set of Gaussian spectral lines that describe the narrow-line component. This spectral decomposition maximizes the broad-line contribution while minimizing the limitations of the 1D optimal extraction.  Analysis of emission line ratios~\citep{1981PASP...93....5B} shows that the dominant excitation mechanism in the central 3--4~arcsec is AGN (Fig.~\ref{fig:bino_maps}), while star formation dominates in the outer parts.

The limited depth of the STIS dataset does not allow detection of the stellar continuum, so these data are used only to model the emission lines. The 0.2~arcsec-wide slit used for STIS observations is much wider than the PSF FWHM, which means that the spectral resolution is primarily defined by the photometric profile of the source. We therefore performed the full spectral fitting of this dataset without convolving the model with the line spread function (LSF). 
We converted fluxes from surface brightness to total flux for point sources using the coefficient provided in the \textit{DIFF2PT} field of the FITS header.

Both the narrow- and broad-line components in the spectroscopic datasets exhibit significant asymmetries and symmetric deviations from pure Gaussian, prompting the inclusion of additional Gaussian-Hermite (GH) terms in the modeled line-of-sight velocity distributions (LOSVDs). For the STIS dataset a proper description of the broad $H\alpha$ shape, including the blue wing, was achieved only with the GH of 6-th degree.  The reduced $\chi^2_{dof}$ for the modeled STIS dataset is 1.035 for a 4-th degree GH, but it decreases to 1.032 for 6-th degree GH  ($\Delta \chi^2 = 3.9$). A 6-th degree GH also adequately describes the outer parts of the broad-line component in the ESI dataset ($\chi^2_{dof}=0.454$ compared to 0.476 for a pure Gaussian; $\Delta\chi^2 = 420$).  The Binospec data are well fit with a 4-th degree GH.   
The resulting values of the highest-order GH coefficients from the STIS ($h_6$) and Binospec ($h_4$) spectra, approximately $0.1$, confirm the pronounced deviations of the broad-line component from a Gaussian.  Although the full $\chi^2$ differences are small, they originate from the small part of the spectrum occupied by the broad lines (tens of pixel vs thousands or tens of thousands pixels for the whole spectral range) and  are statistically significant.

\subsection{Virial black hole mass estimate from the broad-line component profile}

\begin{figure*}[h]
    \centering
    \includegraphics[width=0.95\hsize, trim={0 0.75cm 0 0},clip]{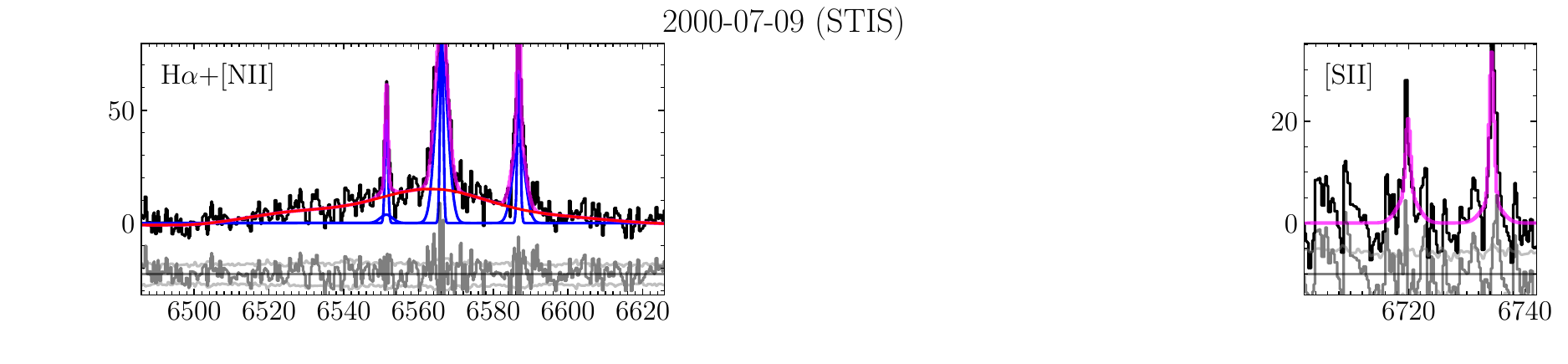}\\
    \includegraphics[width=0.95\hsize, trim={0 0.75cm 0 0},clip]{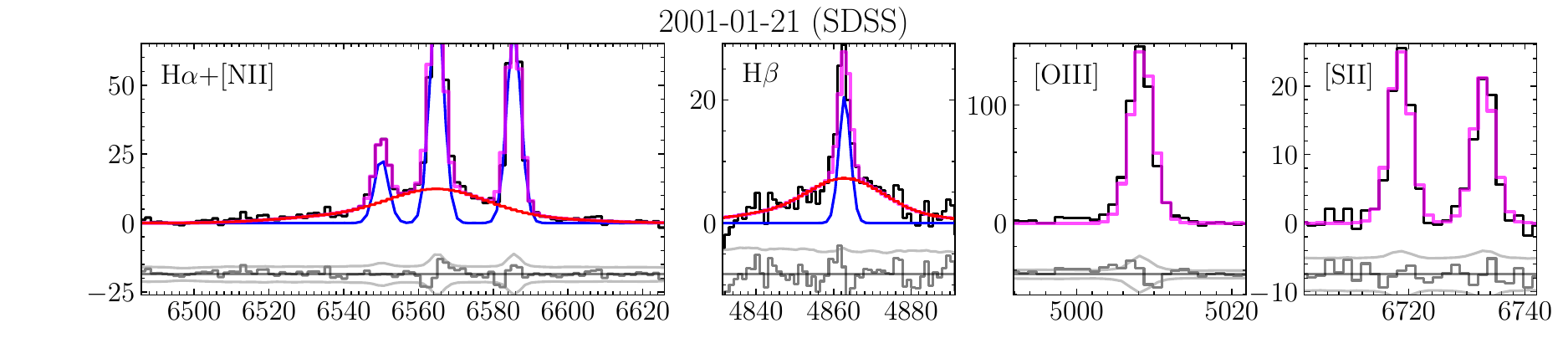}\\
    \includegraphics[width=0.95\hsize, trim={0 0.75cm 0 0},clip]{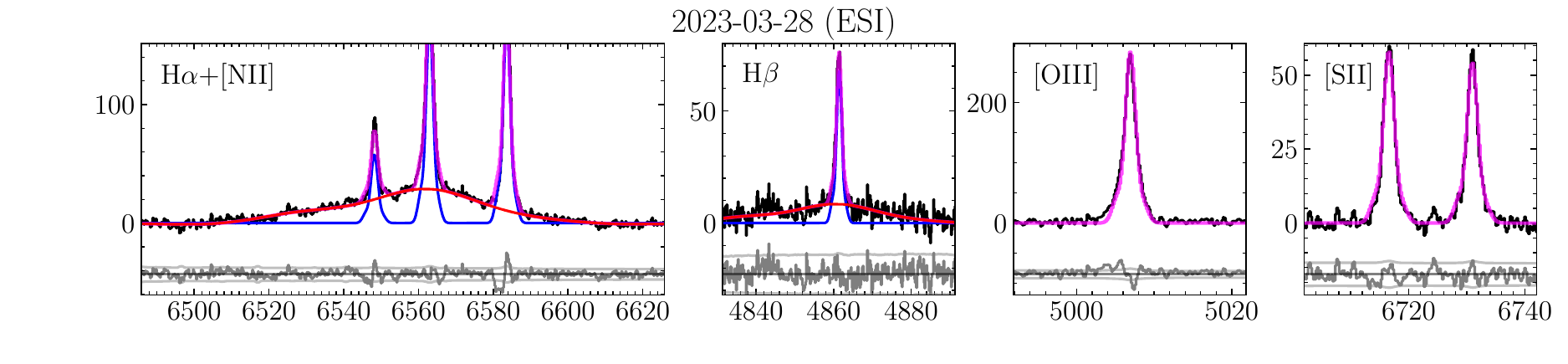}\\
    \includegraphics[width=0.95\hsize]{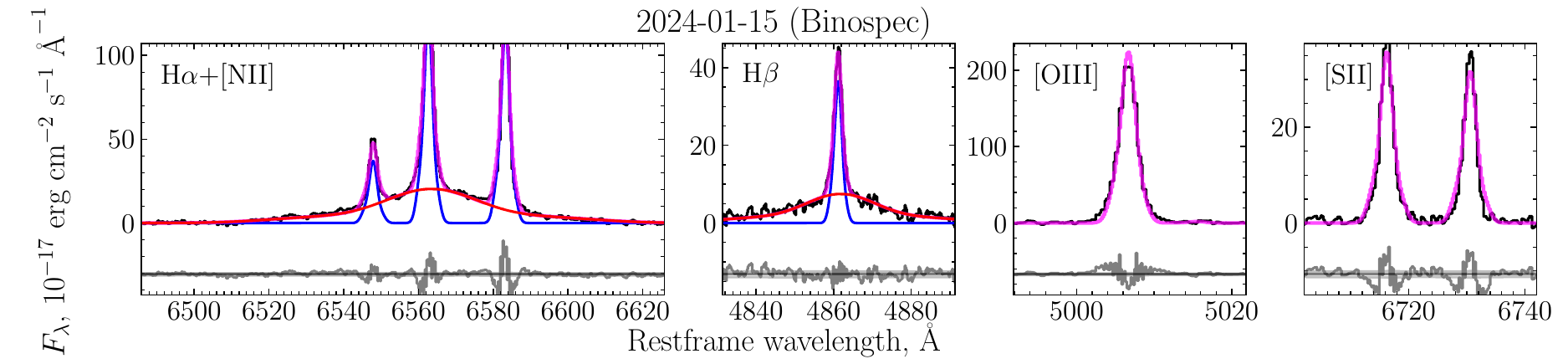}\\
    \caption{Results of the full spectral fitting of the multi-epoch dataset for NGC~3259. In each row: emission line decomposition for the $H\alpha$+[N{\sc ii}], $H\beta$, [O{\sc iii}] and [S{\sc ii}] regions; black - data, blue - narrow line components, red - broad line component, magenta - total model of emission lines, residuals and uncertainties are shown in grey.} \label{fig:spectra}

\end{figure*}

We used the broad-line $\mathrm{H\alpha}$ profile to estimate the black hole mass (Fig.~\ref{fig:spectra}), assuming that the broad line region (BLR) clouds near the black hole are virialized. We used the \citet{2013ApJ...775..116R} relation to derive BH mass from the $H\alpha$  FWHM and luminosity.  Given the significant asymmetry in the emission line profile, we determined the FWHM of the broad-line component by numerically solving the equation that relates the Gaussian-Hermite emission-line profile to half of its maximum flux. 

Our measurements of the emission line fluxes show that the broad $H\alpha$ component varies by a factor of $\sim$2.5. To rule out possible observational systematics we compared $F_{H\alpha, broad}$ with the measurements of $F_{\rm{[OIII]}}$, which is seen mainly in the NLR of the AGN.  Star formation contributes negligibly to [O{\sc iii}] flux. $F_{\rm{[OIII]}}$ emission varies by only 20\%, confirming the strong variability of the central source.

Estimates of the BH mass for different epochs ranged from $1.7$ to $4.1\times10^5 M_{\odot}$ (Table~\ref{tab:blr}). Table~\ref{tab:blr} gives the statistical uncertainties estimated from the model covariance matrix. The high S/N ratios of the spectra result in statistical uncertainties of a few percent. The systematic uncertainties from data reduction, including absolute flux calibration and slit loss correction, are larger. However, the agreement of the four nuclear [O{\sc iii}] flux estimates within 20~\%\ (see above) translates into a $M_{BH}$ difference of only 10~\%\ and suggests the physical variability of the broad H$\alpha$ flux from varying extinction or geometric effects.

The empirical calibration of the virial relation that we use to derive $M_{BH}$ has a 0.3--0.4~dex scatter estimated from comparing it with other methods to estimate $M_{BH}$, including reverberation mapping and stellar dynamics~\citep{2011ApJ...739...28X, 2012ApJ...755..167D}.  This calibration is a major source of uncertainty for the  NGC~3259 $M_{BH}$ estimates. We conclude that the most probable value $M_{BH}$ is between 1.7$\times 10^5 M_{\odot}$ and 4.1$\times 10^5 M_{\odot}$ with the uncertainty of a factor up to 2.5 due to unknown virial factor and uncertain internal extinction. 

\begin{table*}[h]
   \centering
    \caption{Properties of the broad emission line components with statistical uncertainties obtained from the full spectral fitting. Columns: 1. Date of the observation; 2. Instrument name; 3. Velocity offset of the broad-line component with respect to the narrow lines; 4. Full width at half-maximum (FWHM) of the broad-line component; 5-6. Gauss-Hermite coefficients $h_3$ and $h_4$ for broad-line component (for the STIS and ESI datasets we provide $h_5$ and $h_6$ values in a separate row); 7. Flux in broad H$\alpha$ component; 8. Luminosity of broad-line component and 9. Virial estimate of the black hole mass from the H$\alpha$ broad component using relation from \citet{2013ApJ...775..116R}. For the Binospec dataset in the last row we also provide a $M_{BH}$ estimate from the spatial-spectral modeling of broad-line component.}
    \def\arraystretch{1.5}
    \setlength\tabcolsep{1.5pt}
    \begin{tabular}{ c c c c c c c c | c c}
    \hline
    \hline

    Date & Instr. & $v_{broad} - v_{narrow}$ & $FWHM_{broad}$ & $h_3$ & $h_4$ & $10^{17}F_{\mathrm{[O\textsc{iii}]}}$ & $10^{17}F_{\mathrm{H\alpha, broad}}$ & $L_{\mathrm{H\alpha, broad}}$ & $M_{\mathrm{BH}}$\\
     & & km~s$^{-1}$ & km~s$^{-1}$ & & & erg cm$^{-2}$s$^{-1}$ & erg cm$^{-2}$s$^{-1}$ & $10^{37}$erg s$^{-1}$ & $10^3 M_{\odot}$ \\
    \hline
2000-07-09 & STIS  & -228$\pm$41 & 2014$\pm$80 & -0.05$\pm$0.03 & 0.03$\pm$0.03 &  & 749.3$\pm$20.5 & 65.4$\pm$1.8 & 324.9$\pm$26.7\\\cline{5-6}
 & & &  & $h_5$ & $h_6$ & & &\\\cline{5-6}
 & & &  & 0.05$\pm$0.03 & -0.09$\pm$0.03 & & &\\
2001-01-21 & SDSS  & -34$\pm$58 & 1621$\pm$158 & -0.03$\pm$0.05 & 0.08$\pm$0.05 & 561.0$\pm$28.1 & 526.9$\pm$33.7 & 46.0$\pm$2.9 & 176.5$\pm$35.9\\
2023-03-28 & ESI  & -168$\pm$22 & 2017$\pm$58 & -0.09$\pm$0.02 & -0.09$\pm$0.02 & 585.3$\pm$6.5 & 1246.8$\pm$20.2 & 108.8$\pm$1.8 & 412.0$\pm$24.7\\\cline{5-6}
 & & &  & $h_5$ & $h_6$ & & &\\\cline{5-6}
 & & &  & 0.03$\pm$0.02 & -0.07$\pm$0.02 & & &\\
2024-01-15 & Binospec  & 41$\pm$3 & 1762$\pm$12 & 0.00$\pm$-0.00 & 0.12$\pm$0.00 & 626.9$\pm$1.1 & 842.0$\pm$2.5 & 73.4$\pm$0.2 & 260.1$\pm$3.7\\
 & &  & 1867$\pm$22& & & & 756.6$\pm$9.6 & 66.0$\pm$0.8 & 279.0$\pm$7.0\\
    \hline
    \hline
    \end{tabular}
    \label{tab:blr}
\end{table*}

\subsection{Stellar population properties}

The results of the stellar population modeling are presented in Table~\ref{tab:decomp_spop}. Our analysis indicates that the central region of NGC~3259 contains an intermediate-age stellar population (SSP-equivalent age $t_{SSP}=2.4$--$4.6$~Gyr) and a nearly solar metallicity ($-0.15$ to $+0.07$~dex). This age estimate reflects the combined influence of star formation activity in the disk, where young stars are present up to the very center of the galaxy (as seen from the {\it HST} images), and the central component, which is predominantly composed of an older, quiescent stellar population.

The high spectral resolution of the ESI and Binospec (Fig.~\ref{fig:bino_maps}) data enabled precise measurements of the velocity dispersion of the central component in the range $49$--$65$~km~s$^{-1}$. A clearly visible central dip in the velocity dispersion map is likely a signature of a central star cluster that is dynamically colder than the spheroidal component ($\sigma_* \approx 57$~km~s$^{-1}$) of NGC~3259. ESI measurements also suggest that the central star cluster is less metal-rich ($-0.2$~dex) than the spheroid ($-0.05\dots+0.05$~dex).

\begin{table*}[h]
   \centering
    \caption{Properties of the stellar populations (velocity ($v_*$), velocity dispersion ($\sigma_*$), age ($t_{SSP}$) and metallicity ([Z/H])) and narrow emission lines (velocity ($v_{narrow}$) and velocity dispersion ($\sigma_{narrow}$)) inferred from the full spectral fitting. For STIS dataset we provide both kinematical components used for the NLR modeling.}
    \def\arraystretch{1.5}
    \begin{tabular}{ c | c c c c | c c}
    \hline
    \hline
    Instrument & $v_*$ & $\sigma_*$ & $t_{SSP}$ & [Z/H] & $v_{narrow}$ & $\sigma_{narrow}$\\
     & km~s$^{-1}$ & km~s$^{-1}$ & $10^6$ yr & dex & km~s$^{-1}$ & km~s$^{-1}$\\

    \hline
 STIS & & & & & 1681.6$\pm$0.8 & 17.4$\pm$1.0\\
 & & & & & 1680.8$\pm$2.1 & 75.0$\pm$2.8\\
 SDSS & 1699.3$\pm$1.5 & 53.0$\pm$2.5 & 4634$\pm$161 & -0.15$\pm$0.02 & 1681.0$\pm$1.0 & 58.5$\pm$1.4\\
 ESI & 1674.3$\pm$0.8 & 49.9$\pm$0.9 & 4123$\pm$194 & 0.00$\pm$0.03 & 1680.6$\pm$0.4 & 47.4$\pm$0.6\\
 Binospec & 1689.3$\pm$0.2 & 65.2$\pm$0.7 & 2415$\pm$14 & 0.07$\pm$0.01 & 1679.4$\pm$0.1 & 49.0$\pm$0.1\\
    \hline
    \hline
    \end{tabular}
    \label{tab:decomp_spop}
\end{table*}

\subsection{Isophotal analysis of HST F814W dataset}
\label{sec:iso_mod}

To measure structural properties of different components of NGC~3259 we extracted the 1-D light profile using the {\sc isophote.Ellipse}\footnote{\url{https://photutils.readthedocs.io/en/stable/api/photutils.isophote.Ellipse.html}} task from the {\sc Photutils} \textit{Python} package. The isophotes were extracted within ellipses with logarithmically scaled semi-major axes and a step size of 0.1. The extracted 1-D profile was fitted by minimizing the $\chi^2$ metric with the {\sc lmfit} implementation of the Levenberg-Marquardt algorithm\footnote{\url{https://lmfit.github.io/lmfit-py/}}~\citep{2016ascl.soft06014N}. The model includes five components: four Sersic profiles and one PSF. The PSF component represents the unresolved core of the central star cluster and an AGN, the first (inner) Sersic component models an extended part of the nuclear star cluster, the second a central spheroid, the third and the fourth the main disk of the galaxy. For the nuclear star cluster and the main disk, we also tried King~\citep{1962AJ.....67..471K} and exponential profiles, respectively, but these did not provide a better fit. Throughout the modeling process, we convolved our model with the PSF generated using the {\sc TinyTim} tool~\citep{2011SPIE.8127E..0JK} for a G2V spectral energy distribution.

The large number of model parameters may be degenerate, depending on the initial guess.  We used the Markov Chain Monte Carlo Ensemble sample generator \citep[emcee;][]{2010CAMCS...5...65G} implemented in {\sc lmfit}, running 9000 steps to efficiently explore the parameter space. From this procedure, we selected the values that delivered the highest likelihood as the initial guess for the finely-tuned fitting procedure with the Levenberg-Marquardt algorithm.  In Table~\ref{tab:decomp_1d} we provide the results of the decomposition of the 1-D photometric profile based on the {\it HST} dataset in the \textit{F814W} band. 

\begin{figure*}[h]
    \centering
    \includegraphics[width=\hsize]{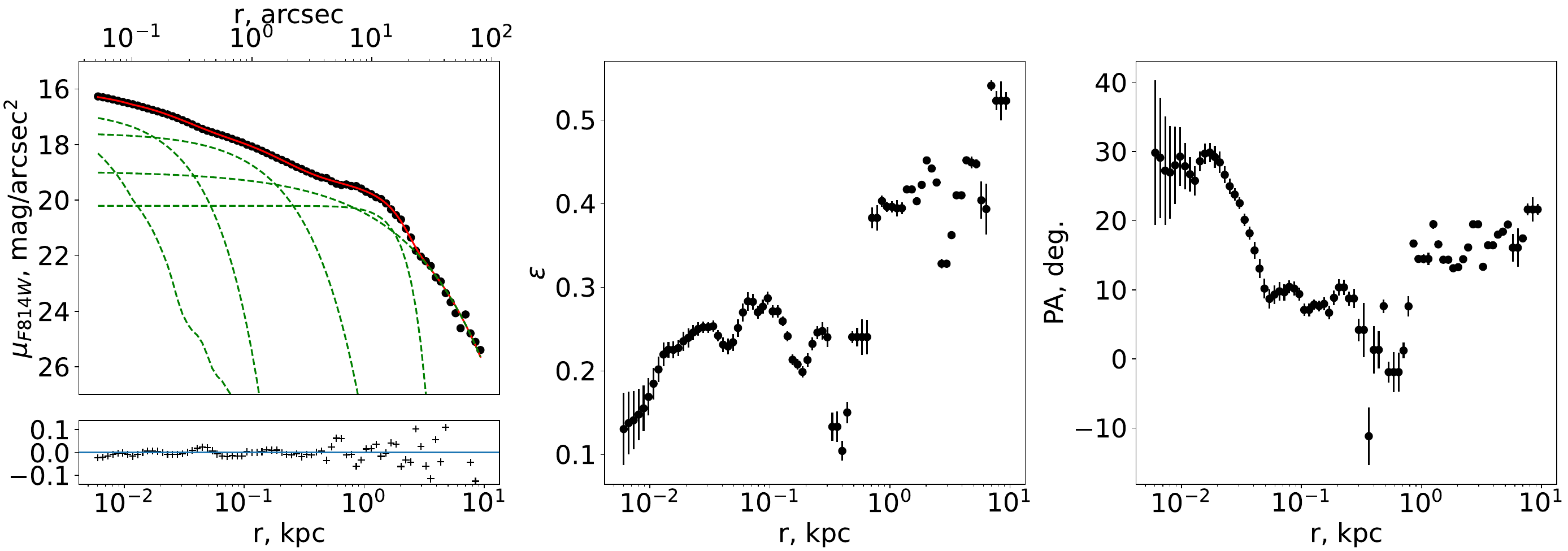}
    \caption{Results of isophotal analysis of {\it HST ACS/WFC} images of NGC~3259 in the {\it F814W} band. Left panel: Surface brightness ($\mu_{F814W}$) profile with the best-fitting model (red line) and its components (green dashed lines); residuals are shown in the bottom sub-panel. Middle panel: ellipticity ($\varepsilon$) of isophotes. Right panel: position angle (PA) of isophotes. \label{fig:acs_decomp}} 
\end{figure*}

The NGC~3259 disk has a complex structure that cannot be adequately described by a single exponential model due to a knee in the surface brightness profile at a radius of approximately 2~kpc (Fig.~\ref{fig:acs_decomp}). We successfully modeled the disk-dominated data at $r > 1~\mathrm{kpc}$ with two S\'ersic components.  The first component describes most of the disk with a nearly exponential profile (S\'ersic index $n=1.49 \pm 0.33$). The second component fits the knee in the surface brightness profile with an effective radius approximately half that of the first component and a S\'ersic index $n=0.31 \pm 0.08$.  The second component declines steeply beyond the effective radius.

The central spheroid, which can represent the bulge, pseudo-bulge, and/or possibly a bar, was fitted with a S\'ersic index $n=0.97 \pm 0.58$, consistent with an exponential profile, and an effective radius of $170 \pm 40~\mathrm{pc}$. The spheroid size is consistent with the smallest observed bulges of galaxies ~\citep[e.g.][]{1992ApJ...399..462B, 2009MNRAS.393.1531G}. It also makes NGC~3259 different from NGC~4395, a bulgeless galaxy \citep{2003ApJ...588L..13F}.


\begin{table*}[h]
   \centering
    \caption{Parameters of the photometric components inferred from the modeling of 1-D surface brightness profile extracted in the isophote analysis of the \textit{F814W} imaging dataset for NGC~3259. Parameters are: 1. Surface brightness at effective radius ($\mu_{R_e}$); 2. Effective radius $R_e$ and 3. Sersic index n. }
    \def\arraystretch{1.5}
    \begin{tabular}{ c c | c c c c c}
    \hline
    \hline
    Parameter & Units & \multicolumn{5}{c}{Components} \\
    \hline
    & & Disk (1) & Disk (2) & Spheroid & Nuc. S.C. & PSF  star star\\ 
    \hline
$\mu_{R_e}$ & mag/arcsec$^2$ & 21.85 $\pm$ 0.33  & 20.58 $\pm$ 0.25  & 19.34 $\pm$ 0.13  & 18.28 $\pm$ 0.44  & 18.00 $\pm$ 0.32 \\
$R_e$ & kpc & 2.66 $\pm$ 0.34  & 1.32 $\pm$ 0.04  & 0.17 $\pm$ 0.04  & 0.03 $\pm$ 0.01 & \\
n & & 1.49 $\pm$ 0.33  & 0.31 $\pm$ 0.08  & 0.97 $\pm$ 0.58  & 0.86 $\pm$ 0.14 & \\

    \hline
    \hline
    \end{tabular}
    \label{tab:decomp_1d}
\end{table*}

\subsection{2D modeling of the nuclear region}

\begin{figure}[h]
    \centering
    \includegraphics[width=\hsize]{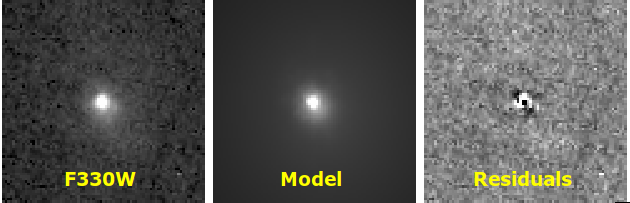}
    \caption{Results of the decomposition of the \textit{F330W HST ACS/HRC} dataset with {\sc galfit}. Each panel measures 1$\times$1~arcsec, North is up, East is left. The double S\'ersic + PSF model is shown.\label{fig:f330w_decomp}} 
\end{figure}

To characterize the nuclear star cluster in NGC~3259, we fit a 2D model to the central 1$\times$1~arcsec$^2$ (160$\times$160~pc$^2$) region containing its nucleus using the \textit{F330W} {\it HST ACS/HRC} dataset (Fig.~\ref{fig:f330w_decomp}). Our model includes three components: (i) a point source, (ii) an extended S\'ersic component for the nuclear star cluster, and (iii) a background component with an exponential profile to fit the diffuse UV emission from the star-forming disk.  The third component has a fixed position and an ellipticity. The final model was convolved with the PSF, generated using the {\sc TinyTim} utility~\citep{2011SPIE.8127E..0JK} at the position of the nucleus of NGC~3259.  A S\'ersic profile is widely used to model resolved nuclear star clusters in nearby galaxies~\citep{2020A&ARv..28....4N, 2023MNRAS.520.4664H}.

Table~\ref{tab:f330w_galfit} summarizes our model of the central regions of NGC~3259. The extended component has a nearly Gaussian shape (S\'ersic index $n=0.6$) and an effective radius $r_e=$14~pc.  The S\'ersic index is lower than typical for nuclear star clusters in nearby galaxies~\citep{2023MNRAS.520.4664H} and the effective radius is larger than typical. $\chi^2_{dof}$ for this model (3.256) is smaller than for a generalized King profile (3.279) with a difference in $\chi^2$ of $\Delta \chi^2=150$, indicating that the S\'ersic profile better describes the data.  $\chi^2_{dof}$ is substantially higher than 1.0 due to the large residuals in the central part of a galaxy dominated by the imperfectly sampled PSF interpolated by the data reduction pipeline (see Fig.~\ref{fig:f330w_decomp}).

\begin{table}[h]
   \centering
    \caption{Properties of central components estimated with Galfit modeling of the \textit{F330W} {\it HST} image. Columns are: 1. Parameter name; 2. Units; 3. Value; Columns 4-5 represent the intrinsic properties of the components estimated from the decomposition. Parameters in this table are: A full magnitude of a component (mag.);  Effective radius ($r_e$); Sersic index ($n$); Axial ratio ($b/a$), Position angle (PA).}
    \def\arraystretch{1.5}
    \begin{tabular}{ c c c | c c}
    \hline
    \hline
    Parameter & Unit & Value & \\
    \hline
\multicolumn{5}{c}{Point Source}\\
\hline
total mag & mag & 20.54$\pm$0.02 & abs. mag. & -11.62$\pm$0.02 \\
\hline
\multicolumn{5}{c}{Extended component}\\
\hline
total mag & mag & 20.41$\pm$0.02 & abs. mag. & -11.74$\pm$0.02 \\
$r_e$ & pix & 4.31$\pm$0.13 & pc & 14.12$\pm$0.43 \\
n  & & 0.625$\pm$0.069 &\\
b/a  & & 0.833$\pm$0.025 &\\
PA & deg & 38.0$\pm$6.4 &\\
    \hline
    \hline
    \end{tabular}
    \label{tab:f330w_galfit}
\end{table}

\subsection{Analysis of X-ray XMM-Newton spectra}

We used the Sherpa package~\citep{2007ASPC..376..543D} to model the coadded X-ray spectrum.  The spectrum cannot be described by a simple power law with dust absorption; $\chi^2_{dof}$ for a single power law fit on the XMM-Newton Survey Science Center service\footnote{\url{http://xmm-catalog.irap.omp.eu/}} is 1.87.  We modeled the spectrum with two components, each consisting of a power law with absorption. One power law describes a central AGN with varying luminosity but constant spectral index and absorption, and the other describes a constant component, softer power law spectrum with absorption. The first absorption component affects only the embedded AGN.  We use the Poisson log-likelihood function (\textit{cstat}) as the goodness-of-fit metric. To ensure sufficient S/N we binned 2006 and 2011 datasets into 20 and 15 channels, respectively.  Table~\ref{tab:xray_spec_prop} summarizes our model for the {\it XMM-Newton} X-ray spectra.

Our model for the soft X-ray component has a power law index $\Gamma = 3.21_{-0.40}^{+0.50}$ with a low absorption column density $N_H = 1.0_{-0.3}^{+0.5}\times10^{21}~\mathrm{cm^{-2}}$. This component is likely associated with extended thermal emission from circumnuclear gas, commonly observed in nearby Seyfert~2 galaxies and Compton-thick highly-obscured AGN~\citep{2022hxga.book...92F}. The outer part of this region in NGC~3259 shows LINER-type excitation in the IFU data (Fig.~\ref{fig:bino_maps}), also common in AGN sub-kpc scale ionization~\citep{2021ApJ...908..155M, 2016ApJ...829...46M}. 

This soft component can also contain some flux contribution from an unresolved population of high-mass X-ray binary stars~\citep{2007A&A...473..783R}. The luminosity in the 2--10 keV band of $F_{\rm{2-10 keV}} = 1.7\times10^{38}~\mathrm{erg~s^{-1}}$ suggests a star formation rate (SFR) of $0.02~\mathrm{M_{\odot}~yr^{-1}}$~\citep{2016ApJ...825....7L}, which we consider as an upper limit estimate. The H$\alpha$ flux is strongly suppressed in this area indicating very low if any current star formation.

Our model for the hard X-ray component has a much higher column density $N_H \sim (4-7) \times 10^{22}~\mathrm{cm^{-2}}$, suggesting that the accretion disk (i.e. the central source) and its corona are obscured by dust. 

Fig.~\ref{fig:xray_spec} shows the best-fitting model for both epochs, with the best-fitting parameters listed in Table~\ref{tab:xray_spec_prop}, and the corresponding \textit{cstat} value of 0.947 (degrees of freedom/datapoints: 24/31). This relatively simple two-component model provides the best fitting quality compared to models with varying $\Gamma$ ($cstat = 0.967$, 23/31), varying $N_{H}$ ($cstat = 0.982$, 23/31), and both varying $\Gamma$ and $N_{H}$ ($cstat = 1.005$, 22/31). Given that \textit{cstat} statistics is distributed similarly to $\chi^2_{dof}$, the differences of \textit{cstat} here are equivalent to $\Delta \chi^2 \sim 1.0$, which is statistically significant in the case of dozens of data points.

A similar absorbed power-law plus a soft component spectrum is observed in NGC~4395 \citep{2000MNRAS.318..879I}. $N_H$ absorption in NGC~4395's hard component varies by almost an order of magnitude between $10^{22}$ and $10^{23}$.

For the 2006 and 2011 datasets, the unabsorbed fluxes for the hard component are $1.14^{+1.16}_{-0.71}$ and $2.32^{+1.47}_{-1.27}$ $10^{-13}~\mathrm{erg~cm^{-2}~s^{-1}}$, respectively. We do not detect X-ray variability with 3$\sigma$ confidence.


\begin{table*}
   \centering
    \caption{Parameters of the best-fitting model for two {\it XMM-Newton} X-ray datasets. Columns are: 1. Date of observation (for the hard component) or component name; 2. Powerlaw slope $\Gamma$; 3. Atomic hydrogen column density on the line of sight ($N_{\rm{H}}$); 4. Unabsorbed flux in 0.2-10 keV band ($F_{\rm{0.2-10~keV}}$); 5. Unabsorbed luminosity in 0.2-10 keV band ($L_{\rm{0.2-10~keV}}$); 6. Eddington ratio for the hard component from unabsorbed fluxes ($\lambda$).}
    \def\arraystretch{1.5}
    \begin{tabular}{ c c c c | c c }
    \hline
    \hline

    Date/Component & $\Gamma$ & $N_{\rm{H}}$ & $F_{\rm{0.2-10~keV}}$ & $L_{\rm{0.2-10~keV}}$ & $\lambda$\\
     &  & $10^{22}$ $cm^{-2}$ & $10^{-14}$erg cm$^{-2}$s$^{-1}$ & $10^{40}$erg s$^{-1}$ & $10^{-2}$\\

    \hline
 Const (Soft) & $3.21_{-0.40}^{+0.50}$ & $0.10_{-0.03}^{+0.05}$ & $6.24_{-3.82}^{+15.79}$ & $0.54_{-0.33}^{+1.38}$ & \\
\hline 
2006-05-06& $0.74_{-0.30}^{+0.32}$& $5.71_{-1.74}^{+2.11}$ & $9.37_{-4.50}^{+9.94}$ & $0.82_{-0.39}^{+0.87}$ & $0.49_{-0.24}^{+0.52}$\\
2011-10-10& &  & $20.07_{-12.32}^{+21.07}$ & $1.75_{-1.07}^{+1.84}$ & $1.05_{-0.64}^{+1.10}$\\

    \hline
    \hline
    \end{tabular}
    \label{tab:xray_spec_prop}
\end{table*}

\section{Discussion}

\subsection{Central black hole in NGC~3259}

In this section, we review the multi-wavelength evidence supporting our classification of NGC~3259 as a low-luminosity LSMBH and not a dusty Seyfert-2 galaxy powered by a considerably more massive BH. 

\subsubsection{Central source obscuration and geometry}

Our analysis of the X-ray data reveals a strongly absorbed hard component, consistent with emission from an obscured AGN accretion disk and hot corona.  The obscuration does not affect the soft component, which may be thermal emission from hot gas in the narrow-line region, or may be emitted by a population of X-ray binaries in the disk or a nuclear star cluster.  

We do not detect a power-law additive continuum in the optical spectra, also consistent with strong extinction of the central source.  However, the optical broad emission lines are not consistent with high extinction. We measure broad-line Balmer decrements of H$\alpha / $H$\beta =3.79\pm0.25$ from the ESI data and $4.61\pm0.71$ from the Binospec data, consistent within 1$\sigma$.  Low flux in the broad $\mathrm{H\beta}$ line reduces the accuracy of these measurements. 

This discrepancy is not unusual, however.  X-ray absorption measured for type-{\sc i} AGN shows a huge scatter with respect to Balmer decrement based extinction estimates, with X-ray derived neutral hydrogen column densities being factors of $\sim$0.5 to $\gtrsim$1000 higher than the Galactic 'average' \citep[e.g.,][]{2001A&A...365...28M, MejiaRestrepo2022}. One explanation for this scatter is that the radiation field of the AGN sublimates dust in its vicinity.  The optically thin ionized gas close to the black hole will not significantly affect the optical line and continuum emission, but atomic and ionized gas will absorb soft X-rays. A second explanation is that many individual BLR and torus clouds orbiting the SMBH are thought to have a comparable angular size to the X-ray corona, such that they can temporarily occult the X-ray corona for days to years, leading to much higher and time-varying line-of-sight column densities \citep[e.g.,][]{Matt2003, Risaliti2005}. 

This extinction discrepancy extends to low-luminosity objects. For example, three low-luminosity AGN ($\log L_{2-10\, \mathrm{keV}} = 40.5 \dots 41.0$) from \citet{2001A&A...365...28M} have factors of 3--5 higher optical extinction than one would expect from the $N_H$ value measured in the X-ray domain. On the other hand, NGC~4395 has a similar $L_X$ and hosts a moderately obscured ($N_H = 3\times10^{22} - 10^{23}$ cm$^{-2}$) AGN, which is highly variable in both luminosity and $N_H$ and is powered by an LSMBH, but its FUV-to-optical absorption is very weak: its FUV spectrum has prominent broad lines and continuum, which enabled the C{\sc iv}-based line--continuum reverberation mapping.

\begin{figure*}[ht!]
    \centering
    \includegraphics[width=0.97\hsize]{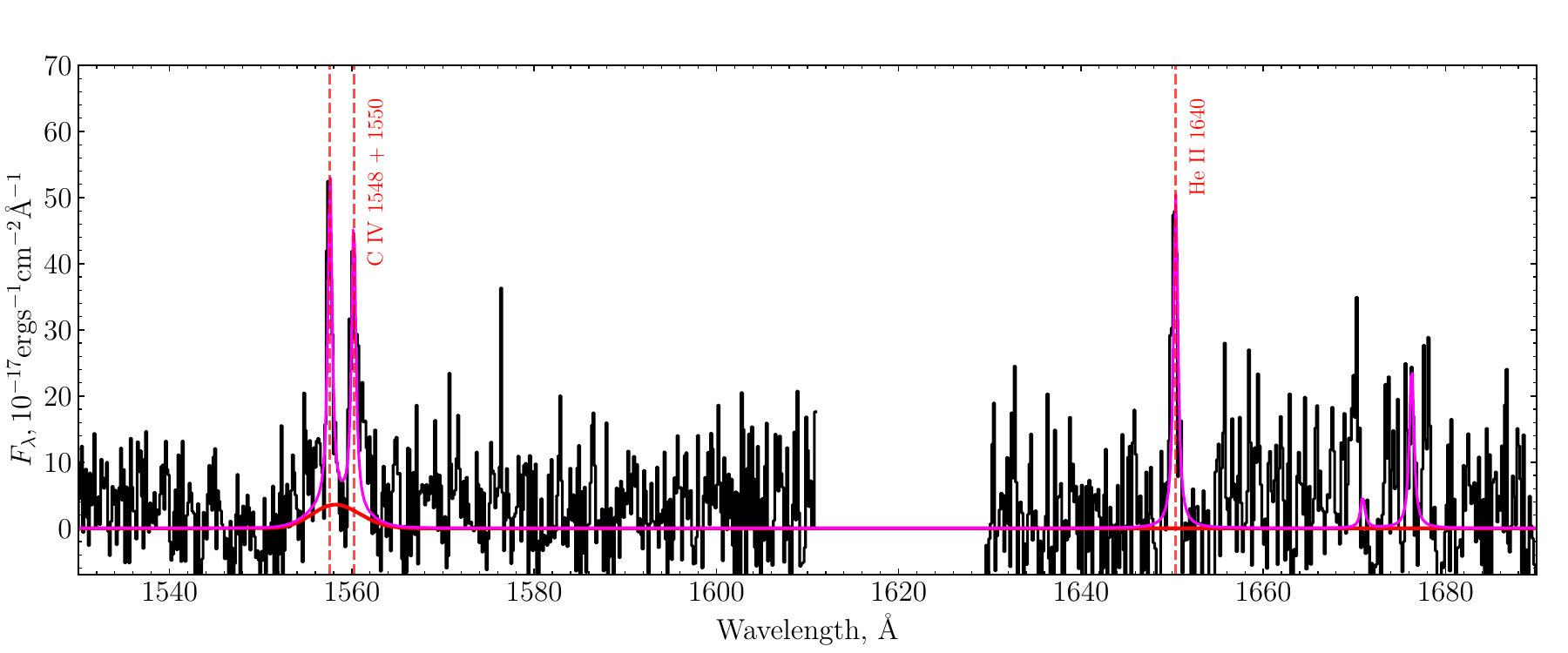}
    \caption{The extracted \textit{HST COS} FUV spectrum of NGC~3259 binned to 0.1~\AA~pix$^{-1}$. We mark the C{\sc iv} and He{\sc ii} lines. We show the best-fitting two-component model of the emission lines in magenta. Narrow lines modeled by the Lorentzian profile. The detection of the C{\sc iv} broad Gaussian component shown in red at the $1.5 \sigma$ level is not statistically significant.} \label{fig:cos_ngc3259}
\end{figure*}

The remaining strong discrepancy between the optical continuum and line extinction could be explained by a nearly edge-on dust torus that heavily obscures the central source but not the BLR.  The extinction in the direction of the BLR can be constrained by the non-detection of the broad FUV C{\sc iv} doublet in the COS spectrum down to the upper limit of $\sim2\times10^{-16}$~erg~cm$^{-2}$~s$^{-1}$ (Fig.~\ref{fig:cos_ngc3259}). 
Keeping in mind that in an unabsorbed AGN the combined flux of the broad components of the C{\sc iv} doublet is a factor $2-3$ higher than H$\alpha$ \citep{2006agna.book.....O}, in NGC~3259 we expect it to be $\sim(2-4)\times10^{-14}$~erg~cm$^{-2}$~s$^{-1}$. Therefore, it is absorbed by about 5--5.5~mag (100--200 times). 

If the FUV-to-optical extinction law is the same as the Galaxy's, we expect $A_V\approx1.5-2$~mag, consistent with the observed broad line Balmer decrement.  A typical Galactic $A_V$-to-$N_H$ ratio of $N_H=(4-5)\times10^{21}$ \citep{2009MNRAS.400.2050G} implies $N_H$ an order of magnitude lower than observed for the hard X-ray component, consistent with our X-ray analysis. If the central source had this moderate optical extinction we would detect optical continuum. The absence of a detectable optical continuum combined with the observed moderate BLR extinction support our interpretation that NGC~3259's active nucleus is powered by a partially obscured LSMBH, partially obscured by a nearly edge-on thin dust torus. 

An intrinsically luminous yet obscured AGN could produce broad lines by scattering from circumnuclear dust clouds \citep[e.g.,][]{Antonucci1985}. In this case the continuum would also be scattered, yielding a type-{\sc i} AGN line-to-continuum ratio and strongly polarized continuum and line emission, contrary to our observations. 


\subsubsection{The nature of broad line variability in NGC~3259}

Detailed analysis of the H$\alpha$ broad-line component reveals time-variable asymmetries in the line profile. If the broad line variability is due to variations in the intrinsic AGN luminosity, we should see an anti-correlation between the observed flux and width of the broad lines. However, in NGC~3259 we see that the broad-line flux increase is accompanied by an increased width and asymmetry in the line (ESI and STIS spectra; see Table~\ref{tab:blr}).

Alternatively, the observed variable flux and asymmetries in the broad $\mathrm{H\alpha}$ could be attributed to variable dust obscuration of the BLR. At different epochs, dust clouds within the torus obscure different regions of the BLR, although much of the BLR remains visible. The broad $\mathrm{H\alpha}$ emitting clouds in the central part of the BLR are more significantly affected by obscuration than those in the outer regions, leading to more pronounced changes in the higher velocity wings of the broad lines. 

In the case of a nearly edge-on orientation of a thin torus, variation in extinction could explain the observed broad $\mathrm{H\alpha}$ variability in flux and shape. If the torus has a small scale-height, the central source may experience significant obscuration, while the BLR remains comparatively less affected. This scenario explains both the variability and the asymmetry of the broad $\mathrm{H\alpha}$ without invoking intrinsic BLR variability. However, it requires an unusual torus geometry, which makes it look more like a thin dusty disk.



The orbital timescales of gas clouds in the dusty torus and the BLR can be estimated using Kepler’s laws, assuming the clouds are in circular orbits dominated by the SMBH gravitational potential. These timescales provide crucial insights into the structure and dynamics of the AGN environment, influencing variability studies and reverberation mapping techniques \citep{1993PASP..105..247P,2001sac..conf....3P}.

The torus sublimation radius is set by the balance between dust grain sublimation and the AGN radiation field. Empirically, the dust sublimation radius scales with the UV luminosity as $r_{sub} \approx 0.36\,(L_{\rm UV, 45})^{0.5}$\,pc \citep[e.g.,][]{2010A&A...515A..23H}, where $L_{\rm UV,45}$ is the UV luminosity in units of 
$10^{45}$ erg\,s$^{-1}$. We adopt $L_{\rm UV}=20\,L_{0.2-10\,keV}$ \citep[e.g.,][]{2012MNRAS.425..623L, 2024A&A...691A.203G} with a value of $L_{0.2-10\,keV}=1.5\times10^{40}$~erg\,s$^{-1}$ from Table~\ref{tab:xray_spec_prop}. This yields 
$r_{sub}\approx0.006$~pc~$\approx1.9\times10^{16}$\,cm. Assuming a Keplerian orbit for our best-fit black hole mass, this implies an orbital period of $\ga$100~yr. This long timescale suggests that the torus structure, which is presumably the primary obscurer for the BLR, should remain relatively stable over the observational timescales of the broad-line variations we observe. Thus, line-of-sight obscuration is an unlikely cause of the variability.

Another plausible scenario is an asymmetric distribution of BLR clouds orbiting the black hole. For the BLR clouds, the virial radius can be estimated at $r_{BLR}\approx G M_{BH} / f v_{BLR}^2$, where
we adopt the maximum observed value of 2000~km~s$^{-1}$ for $v_{BLR}$ and a virial factor $f\approx4$ to account for geometric projection effects from reverberation mapping studies \citep[e.g.,][]{2015PASP..127...67B}. This yields $r_{BLR}\approx0.8-1.6$ light-days and an orbital period of $\approx$60--120~days \citep[e.g.,][]{2005ApJ...629...61K}. Intriguingly, this estimate is below the shortest timescales on which we observe variations in the broad H$\alpha$, and thus remain a possibility; this however might imply that the BLR is not fully virialized, and our BH mass estimates may be biased. One difficulty with the asymmetric BLR interpretation is that we only observe a blue-sided or a symmetric profile, and never a red-sided wing; with only four available spectroscopic epochs, this could still be a chance alignment.

Finally, the occasional blue wing excess and variability might be associated with frequent outflow events. Blueshifted asymmetric wings in broad emission lines have been widely interpreted as signatures of outflowing gas \citep[e.g.,][]{2003ARA&A..41..117C,2005ARA&A..43..769V}. The key reasoning is that if a component of the BLR is moving radially outward from the AGN, the near-side (approaching the observer) produces a blueshifted broad emission, while the receding side may be obscured by the torus or the accretion disk \citep{1995ApJ...454L.105M}. This creates a net blue asymmetry in the line profile.

Further confirmation of the above scenarios can come from velocity-resolved reverberation mapping \citep[e.g.,][]{2018ApJ...869..142D} or detecting associated UV/X-ray outflows.

\subsubsection{Properties of the narrow line region}

Baldwin-Phillips-Terlevich~\citep[BPT;][]{1981PASP...93....5B} classification in each spaxel (Fig.~\ref{fig:bino_maps}) of the IFU dataset clearly shows that the area of the AGN excitation mechanism has a disky morphology with the same position angle and inclination as the main disk of NGC~3259. This points to the dominating excitation by the AGN radiation in the thin disk, where the rotating gas is ionized. Such a shape of NLR also suggests that there are no other significant gas reservoirs outside of the galaxy disk close to the nucleus (e.g. extraplanar gas). 

We observe a 300--900 pc region around the nucleus with AGN excitation.  The size of this region is consistent with the isophotal radius $R_{int} = 833 \pm 310$~pc derived from the correlation of NLR ($R_{int}$) and $L_{\rm{[OIII]}}$~\citep{2022ApJS..260...31Z}. 

The gas velocity dispersion is larger in the NLR area (Fig.~\ref{fig:bino_maps}, second panel in the second row). The increase possibly indicates a biconical outflow from the active nucleus, frequently observed in Seyfert galaxies. Another possibility is that the gas in the inner region is excited by shocks originating from the bar. However, the narrow He{\sc ii} ($\lambda=4686$~\AA) line seen in the circumnuclear region of NGC~3259 without strong star formation suggests the presence of an AGN.

\subsection{The inner bar and the central spheroid}
In the Third Reference Catalogue of Bright Galaxies \citep[RC3;][]{RC31991} NGC~3259 is classified as an intermediate-bar galaxy, SAB(rs)bc.
The S\'ersic index of the central spheroid is between 1 and 2, consistent with a pseudo-bulge~\citep{2004ARA&A..42..603K}, suggesting that this component was not assembled through mergers.  The central component of NGC~3259 has lower ellipticity (0.2--0.3) than the disk (0.4--0.5), suggesting a spheroidal component or a bar. However, the velocity dispersion does not deviate significantly from the Faber--Jackson (\citeyear{1976ApJ...204..668F}) relation for bulges. 


The H$\alpha$ surface brightness displayed in the Binospec IFU map (Fig.~\ref{fig:bino_maps}) decreases by an order of magnitude inside the central 5~arcsec (650~pc). This probably reflects the intrinsic decrease of the H$\alpha$ surface brightness and not dust absorption because the Balmer decrement remains between 2.7 and 2.9.  Neighboring emission lines that would be similarly absorbed by dust, [N{\sc ii}] and [S{\sc ii}], do not show a similar decrease (Fig.~\ref{fig:bino_maps}).  Galaxies with bars that funnel gas inside the inner ring towards the galaxy center~\citep{1993A&A...271..391C, 2007A&A...474...43V, 2020A&A...644A..79G} show similar behavior. The gas and stellar velocity field isovels are S-shaped and inclined, also typical of barred galaxies~\citep{Lindblad1996,2015A&A...576A.102A, Toky2015}.

\begin{figure}
    \centering
    \includegraphics[width=\hsize]{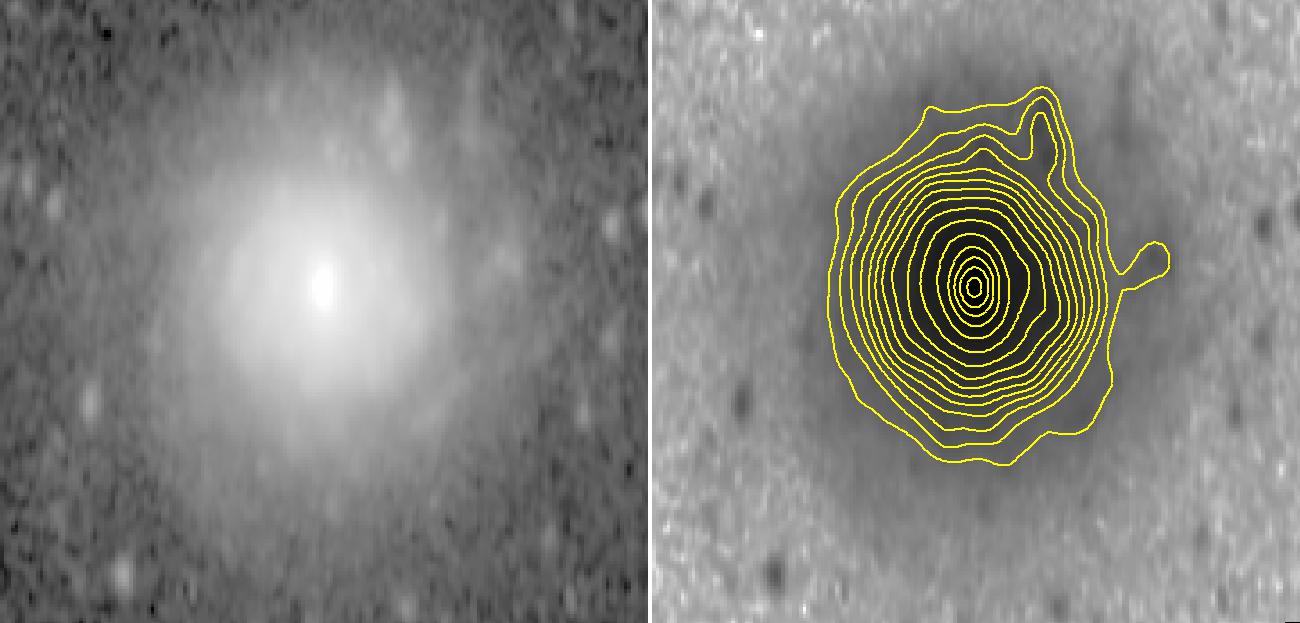}
    \caption{Left: Deprojected {\it Spitzer IRAC1} image for NGC~3259. Right: The same with inverted color map and overlaid contours from 10 to 170 times the background level.} \label{fig:spitzer_deproj_ngc3259}

\end{figure}

However, despite strong kinematic evidence of a bar, optical to near-infrared images fail to provide the clear morphological signature of a bar. Only the \textit{HST NICMOS2} dataset in the \textit{F160W} band subtly indicates an elongated structure on scales of $2$--$3$~arcsec from the center.  Isophotal analysis of this region shows an increase in ellipticity from $0.1$ to $0.3$, which subsequently returns to $0.1$.  In some situations, S-shaped isovelocity lines are less ambiguous evidence of a bar than the photometric analysis. Fig.~\ref{fig:spitzer_deproj_ngc3259} shows the deprojected {\it Spitzer} image of NGC~3259, with an extended component typical of bars based on the position angle PA=16.5 deg. and inclination of $i$=61.1 deg. according to HyperLeda database\footnote{\url{http://atlas.obs-hp.fr/hyperleda/ledacat.cgi?o=NGC3259}}. However, the image deprojection may suffer from an artifact where a slightly elliptical spheroidal component is stretched along the minor axis. Bar identification in inclined galaxies via photometric analysis is problematical for bar major axis orientations roughly aligned with the line of sight.

Observations and simulations show that the dominant formation scenario of pseudo-bulges is secular evolution~\citep{2004ARA&A..42..603K, 2013MNRAS.428..718O} mainly driven by a bar~\citep{Combes1990,2016ASSL..418..413C}. After funneling the gas towards the galaxy center, gravity torques can stall the gas at the inner Lindblad resonance, preventing it from reaching the AGN~\citep{Buta1996,2015MNRAS.454.3641F}. Growth of the central black hole and pseudo-bulge can be suppressed despite the presence of a massive gas reservoir.  NGC~3259 has such a large gas reservoir with a mass of $M_{\rm{HI}} = 6.34\times10^9 M_{\odot}$~\citep{2014A&A...567A..68D}.

\subsection{The lack of the black hole -- spheroid coevolution}

\begin{figure}
    \centering
    \includegraphics[width=\hsize]{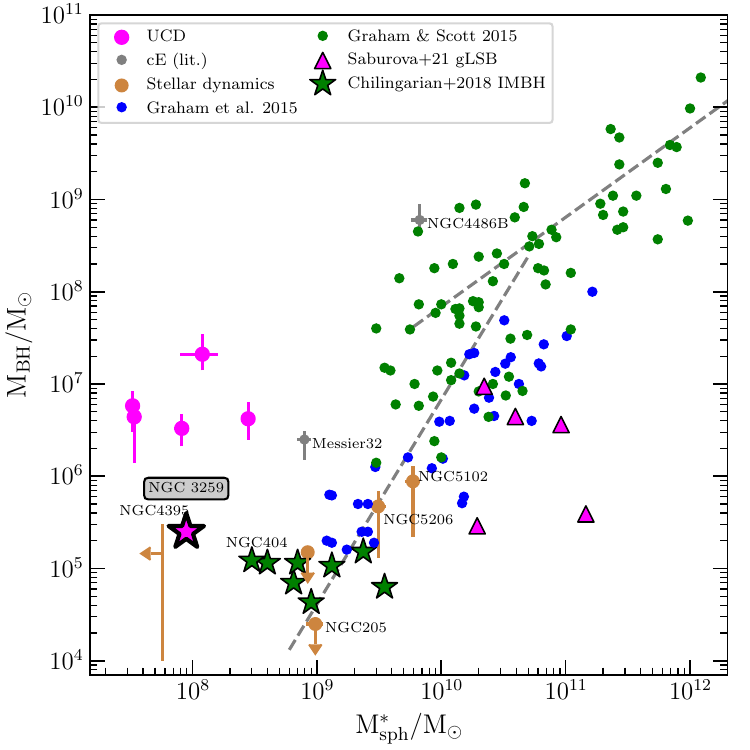}
    \caption{Position (magenta star) of NGC~3259 on the $M^*_{sph} - M_{BH}$ scaling relation. Green stars -- bona-fide sample of IMBH from \citet{2018ApJ...863....1C}. Magenta triangles: giant low-surface brightness galaxies (gLSB) from \citet{2021MNRAS.503..830S} with bulges formed mostly via in-situ processes so the stellar masses of their central spheroids do not correlate with BH masses.} \label{fig:mbulgembh}
\end{figure}

NGC3259's estimated spheroid stellar mass of $M_{sph}^* = 8.9\pm1.2\times 10^7~{\rm M_{\odot}}$ is about an order of magnitude smaller than predicted by the scaling relation between $M_{\mathrm{BH}}$ and $M^{*}_{\mathrm{sph}}$\citep{2015ApJ...798...54G}.  NGC~3259 is nearly a $3\sigma$ outlier from the relation (Fig. \ref{fig:mbulgembh}), a deviation also commonly observed in pseudo-bulges~\citep{2013ARA&A..51..511K}. Since the scaling relation is primarily established for classical bulges, which grow with their central BHs through galaxy mergers~\citep{2013ARA&A..51..511K}, we infer that mergers are not the principal growth mechanism for the central BH in NGC~3259. Instead, gas accretion appears to be the dominant growth channel.
A close analog of NGC 3259, NGC 4395, hosting an intermediate-mass black hole, demonstrates a prominent bar, however, it does not have a bulge, consistent with $M^*_{sph}$-$M_{BH}$ scaling relation, similarly to NGC 3259~\citep{2003ApJ...588L..13F}.

\subsection{Implications for the population of IMBHs in the local Universe}
Identification of low black hole mass AGNs ($M_{BH} < 10^6 M_{\odot}$) is limited by the broad-line component flux and width. Because NGC~3259 is relatively nearby, the detection of broad emission lines in its spectrum is possible. The correlation between the broad line and the X-ray luminosities~\citep{2015ApJ...815....1U} clearly confirms the presence of an AGN. However, in general, low-mass AGNs with low accretion rates are difficult to detect with current spectroscopic and X-ray surveys.

\subsubsection{Simulations of detectability of NGC~3259-like objects}

To explore the detectability of NGC~3259-like objects in spectroscopic surveys such as SDSS and DESI, we performed a set of simulations using the Binospec data cube as input. We applied the same modeling to the simulated spectra using the NBursts spectral fitting package following the strategy for identifying low-mass AGNs by searching for objects with a broad emission line component in their spectra \citep{2005ApJ...630..122G, 2007ApJ...670...92G, 2018ApJ...863....1C}.  We simulated spectra for an NGC~3259-like object at different distances and continuum signal-to-noise (S/N) levels. To mimic observations with a specific instrument at a given distance, we extracted spectra from the Binospec IFU cube within an aperture that matched such an observation. The extracted spectra were then convolved with a Gaussian, making the output spectrum resolving power matched to the simulated instrument by adjusting for the differences between the LSF of SDSS or DESI and Binospec. Finally, the degraded spectrum was resampled to the wavelength scale of the survey, and Gaussian noise was added to match the desired S/N level.  We generated spectra with S/N ranging from 5 to 100 in steps of 5 and angular diameter distances from 5 to 100~Mpc in steps of 5~Mpc. For each combination of these two parameters, we generated 100 spectra with independent noise realizations.

We analyzed the simulated spectra using the {\sc NBursts} package. The stellar continuum was modeled with X-Shooter stellar population models, while Balmer emission lines were represented with a narrow and a broad component. We applied a 9th degree multiplicative stellar continuum and a 2nd degree additive AGN continuum.
Following \citet{2018ApJ...863....1C}, AGN detection is based on the difference in $\chi^2$ values between models with and without a broad-line component to determine where adding a broad component does not significantly improve the agreement between the model and the data. To calculate $\chi^2$ for the narrow-line-only scenario, we also ran the modeling without the second kinematic component for the broad emission lines, using the same procedure as for the case of the broad-line component.

Each simulated spectrum includes a broad-line component, but unreliable broad-line detections are filtered using criteria similar to those in \citet{2018ApJ...863....1C}: $M_{BH} / \Delta M_{BH} > 3$ and $\chi^2 - \chi^2_{broad} > 400$ while number of d.f. was 1546 for modeling with broad-line component and 1550 for modeling without it. Additionally, we limited the velocity dispersion of the broad component to exclude cases with significant model mismatch.

\begin{figure*}[ht!]
    \centering
    \includegraphics[width=0.45\hsize]{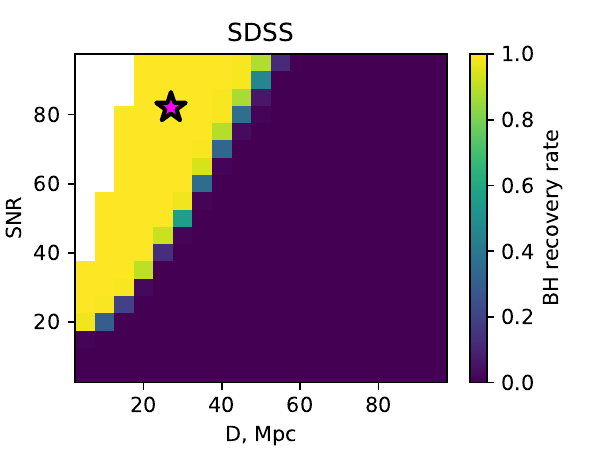}
    \includegraphics[width=0.45\hsize]{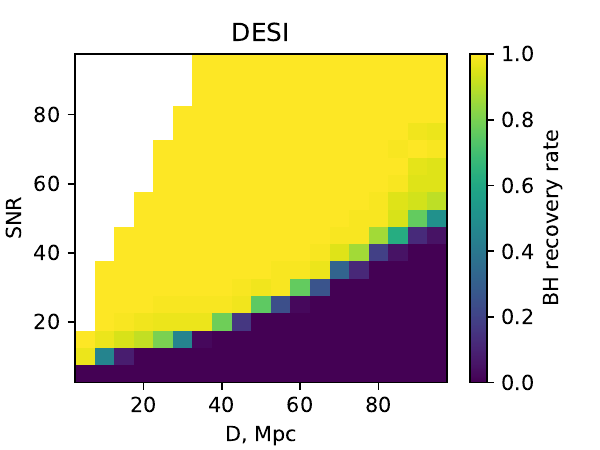}
    \caption{Detectability of broad-line components simulated using the Binospec IFU NGC~3259 datacube  for SDSS (left) and DESI (right). In each bin, the detection probability is color coded.  White cells show the parameter space not simulated due to the limited datacube S/N. A pink star depicts the position corresponding to the NGC~3259 SDSS spectrum.} \label{fig:bh_det_ngc3259}
\end{figure*}

Fig.~\ref{fig:bh_det_ngc3259} presents the detectability of the AGN in NGC~3259 as a function of distance and S/N. For SDSS, there is a strong distance threshold at 35 Mpc where even a large increase in S/N cannot significantly improve the chances of detecting the broad component. The currently observed SDSS spectrum falls within the region of nearly 100\% detectability, validating the simulations. However, these results suggest that NGC~3259 is likely a unique object in the entire SDSS sample, as it is very close to the detectability limit.

Detection of objects like NGC~3259 is expected to be more efficient with the ongoing DESI survey. At the same signal-to-noise (S/N) level, the limiting distance for DESI observations is 3.5–-4 times larger than for SDSS, expanding the volume with high detection probability by up to 64 times. This improvement enables the detection of a larger population of LSMBHs with low accretion rates. Two factors dominate this improvement: (i) a smaller aperture size (1.5~arcsec vs. 3.0~arcsec for SDSS) reduces contribution from the stellar population and (ii) higher spectral resolution and better wavelength sampling disentangle broad components from narrow emission lines and stellar population absorption lines.

\subsubsection{Comparison of NGC~3259 to a rapidly accreting IMBH}

The simulations for NGC~3259 show that the broad-line component detection probability strongly depends on its amplitude relative to the contribution of the stellar population. At larger distances, resolving the stellar population is more difficult, lowering the contrast of the broad-line component.  

We also study how detectability depends on the accretion rate by performing similar simulations for an AGN from a sample of confirmed intermediate-mass black holes~\citep{2018ApJ...863....1C} recently observed with the Binospec-IFU: J110731.23+134712.8. This black hole has a mass of $M_{BH} = (0.71-1.20) \times 10^5 M_{\odot}$, which is 2 to 3 times smaller than NGC~3259.  Its broad-line component is 3–4 times narrower than NGC 3259's but its X-ray luminosity is 100 times higher, suggesting an Eddington ratio two orders of magnitude higher than for NGC~3259.

\begin{figure*}[ht!]
    \centering
    \includegraphics[width=0.45\hsize]{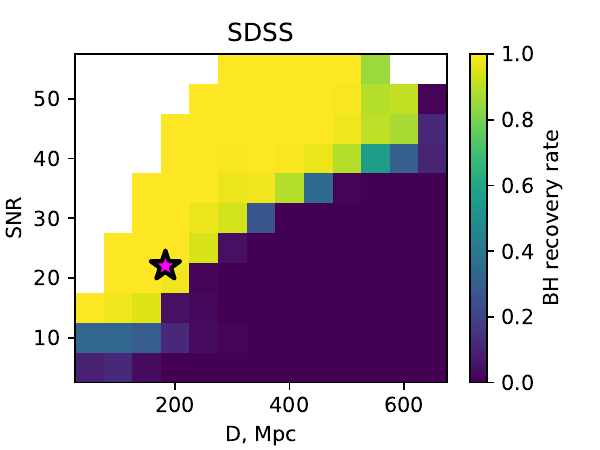}
    \includegraphics[width=0.45\hsize]{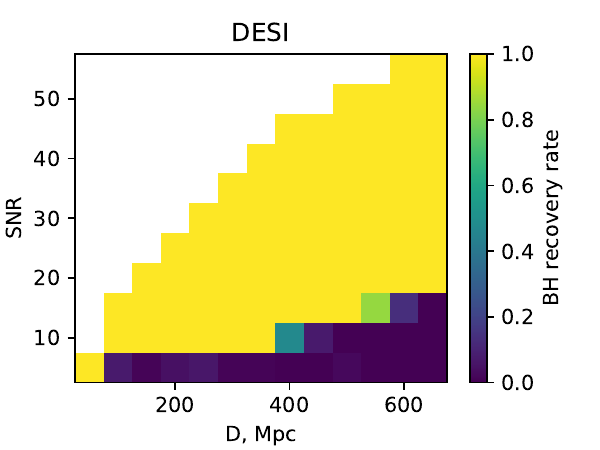}
    \caption{The same as Fig.~\ref{fig:bh_det_ngc3259} but for J110731.23+134712.8.} \label{fig:bh_det_j1107}
\end{figure*}

We present these simulations in Fig.~\ref{fig:bh_det_j1107}. This source is detected in SDSS at a distance 10 times the detection limit for NGC~3259 at similar S/N due to the 10 times higher amplitude and contrast of the broad-line component. At distances $>$400 Mpc, the optical fiber covers the entire galaxy, and for a fixed S/N the shape of the spectrum is independent of the distance. For S/N$\sim$40 J110731.23+134712.8 is detectable to  D$>$400 Mpc, a volume $\sim$4000 times larger than that for NGC 3259.  We therefore miss a large population of LSMBHs accreting at very low Eddington ratios, but we easily detect IMBHs with high accretion rates with SDSS sensitivity. 

This situation drastically improves with spectroscopic surveys like DESI~\citep{2016arXiv161100036D} and 4MOST~\citep{2012SPIE.8446E..0TD}, which both use smaller optical fibers and offer higher spectral resolution than SDSS. With DESI, J110731.23+134712.8-like objects will be detected up to $10^3$ Mpc at S/N=20, which will allow studies of samples of these AGNs in a cosmological context. Nevertheless, building AGN samples confirmed by X-ray observations will be limited by the availability of appropriately sensitive X-ray instruments and surveys covering a large fraction of the sky.

\subsubsection{Limitations of X-ray observations}

The eRASS1~\citep{2024A&A...682A..34M} soft X-ray survey from the first (2019--2020) full eROSITA scan of the sky has a typical depth of $F_{X} \sim 5\times 10^{-13}~\mathrm{erg~cm^{-2}~s^{-1}}$, three times shallower than needed to detect NGC~3259. The expected depth of the eRASS4 survey completed before the end of eROSITA operations in 2022, will marginally detect nearby systems like NGC~3259. However, if our interpretation of a nearly edge-on dusty torus is valid we should expect many systems with similar intrinsic X-ray luminosities but without absorption. These sources will be detected in eRASS4 to $D=60-80$~Mpc and will have a variable optical continuum detectable in time-domain surveys.

AGN candidates can be observed individually with the {\it XMM-Newton} and {\it Chandra} observatories, with exposure times adjusted to reach expected flux levels. The Chandra Source Catalog Release 2 Series reaches the depth of the NGC~3259 detection for almost all observations~\citep{2024ApJS..274...22E}. However, if NGC~3259 was at a distance of 150 Mpc, it would only be detected in half of the Chandra archival datasets according to \citet{2024ApJS..274...22E}. More than half of the exposures in the fourth XMM-Newton serendipitous source catalog~\citep{2020A&A...641A.136W} exceed 20~ks, allowing detection of NGC~3259-like objects to 250~Mpc.

Future X-ray missions will detect more low-accretion LSMBHs. The Advanced X-Ray Imaging Satellite~\citep[AXIS;][]{2023SPIE12678E..1ER} mission, planned for the 2030's, will have ten times higher sensitivity than Chandra. The Advanced Telescope for High Energy Astrophysics~\citep[Athena;][]{2017AN....338..153B} will have ten times larger effective area than XMM-Newton and a larger field of view.

\subsubsection{Limitations on the detection of the AGN continuum variability}

An AGN can also be identified by variable optical continuum using the Zwicky Transient Facility~\citep[ZTF;][]{2019PASP..131a8002B}, the Asteroid Terrestrial-impact Last Alert System~\citep[ATLAS;][]{2018PASP..130f4505T} and the upcoming Legacy Survey of Space and Time~\citep[LSST;][]{2009arXiv0912.0201L}.  These surveys will allow us to identify large, relatively complete samples of AGN, including low-mass AGNs: IMBHs and LSMBHs~\citep{2024ASPC..535..283D}.

Detection of the NGC~3259 continuum variability is prevented by strong obscuration of the central source. The NGC~3259 AGN continuum is less than 1\%\ of the $g$-band flux in the SDSS aperture of 3~arcsec, which translates to $<$0.02\%\ of the integrated galaxy flux. This value is below the level of variability of most low-luminosity AGN: 84\% of the objects from the weak type-I AGN sample from \citet{2023MNRAS.518.1531L} exhibit $g$-band variability above 0.4\%.  Reaching even 1\%\ sensitivity for a $>$19-th magnitude point source superposed on a bright host galaxy is challenging for ground-based time-domain surveys. However, unabsorbed AGN powered by low-Eddington LSMBHs and IMBHs similar to NGC~3259 but with different dust torus orientation should be easily detectable.

\section{Conclusions}

We study NGC~3259 as an example of a LSMBH with $M_{BH} < 10^6 M_{\odot}$ accreting at the low Eddington ratio with the corresponding low AGN luminosity in the X-ray and in broad H$\alpha$. While identification of similar galaxies with SDSS is difficult, DESI and future 4MOST observations will enlarge the sample by orders of magnitude.

Our study of NGC~3259's X-ray emission finds that X-rays from the AGN are obscured by a factor of $\sim$100 in the 0.2--2 keV band, which limits AGN detectability in X-rays and complicates separating the contribution from the stellar population, primarily high mass X-ray binaries. 

Because NGC~3259 is located in a sparse galaxy group, it likely did not have a rich merger history. The presence of a pseudo-bulge instead of a classical bulge supports this conclusion. The central spheroid probably evolves separately from the black hole, in contrast with the coevolution picture, where both grow simultaneously through mergers. Indeed, NGC~3259 departs by almost an order of magnitude from the $M^*_{sph}$-$M_{BH}$ scaling relation, in contrast with more massive SMBHs found in massive bulges and IMBHs that are accreting at high Eddington ratios.


\begin{acknowledgements}
Observations reported here were obtained at the MMT Observatory, a joint facility of the Smithsonian Institution and the University of Arizona. Kirill A. Grishin acknowledges support from ANR-24-CE31-2896. Igor Chilingarian’s research is supported by the SAO Telescope Data Center. He also acknowledges support from the NASA ADAP-22-0102 grant. Franz E. Bauer acknowledges support from ANID-Chile BASAL CATA FB210003, FONDECYT Regular 1241005, and Millennium Science Initiative, AIM23-0001. We thank G.~Fabbiano and M.~Elvis for fruitful discussions on the topic. Some of the data presented herein were obtained at Keck Observatory, which is a private 501(c)3 non-profit organization operated as a scientific partnership among the California Institute of Technology, the University of California, and the National Aeronautics and Space Administration. The Observatory was made possible by the generous financial support of the W. M. Keck Foundation. 
Funding for the SDSS and SDSS-II has been provided by the Alfred P. Sloan Foundation, the Participating Institutions, the National Science Foundation, the U.S. Department of Energy, the National Aeronautics and Space Administration, the Japanese Monbukagakusho, the Max Planck Society, and the Higher Education Funding Council for England. The SDSS Web Site is \url{http://www.sdss.org/}.
The SDSS is managed by the Astrophysical Research Consortium for the Participating Institutions. The Participating Institutions are the American Museum of Natural History, Astrophysical Institute Potsdam, University of Basel, University of Cambridge, Case Western Reserve University, University of Chicago, Drexel University, Fermilab, the Institute for Advanced Study, the Japan Participation Group, Johns Hopkins University, the Joint Institute for Nuclear Astrophysics, the Kavli Institute for Particle Astrophysics and Cosmology, the Korean Scientist Group, the Chinese Academy of Sciences (LAMOST), Los Alamos National Laboratory, the Max-Planck-Institute for Astronomy (MPIA), the Max-Planck-Institute for Astrophysics (MPA), New Mexico State University, Ohio State University, University of Pittsburgh, University of Portsmouth, Princeton University, the United States Naval Observatory, and the University of Washington.
Based on observations made with the NASA/ESA Hubble Space Telescope, and obtained from the Hubble Legacy Archive, which is a collaboration between the Space Telescope Science Institute (STScI/NASA), the Space Telescope European Coordinating Facility (ST-ECF/ESA) and the Canadian Astronomy Data Centre (CADC/NRC/CSA).
This research is based on observations made with the NASA/ESA Hubble Space Telescope obtained from the Space Telescope Science Institute, which is operated by the Association of Universities for Research in Astronomy, Inc., under NASA contract NAS 5–26555. These observations are associated with programs 8228, 9395, 12557.

\end{acknowledgements}
 

\bibliographystyle{aa}
\bibliography{references.bib}

\begin{appendix}
\section{Integration of a Gaussian profile in hexagonal IFU spaxels}
\label{app:green}
To precisely evaluate the 3D point-source broad line component in the Binospec datacube we integrate the the model PSF flux within each spaxel.  Only this approach provides accurate results for undersampled datasets~\citep{2025arXiv250117163C}. However, two-dimensional numerical integration is computationally expensive.

\begin{figure}[ht!]
    \centering
    \includegraphics[width=\hsize]{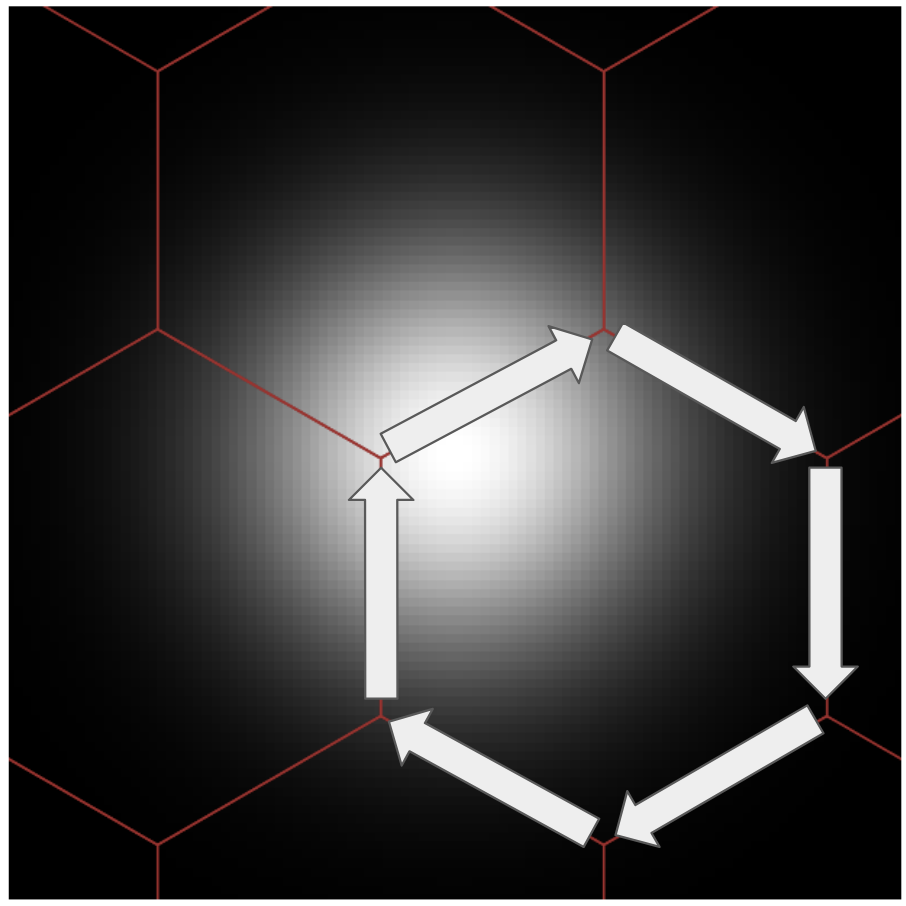}
    \caption{An example of integration over a hexagonal contour overlaid on a 2D Gaussian profile.} \label{fig:cont_integ}
\end{figure}

To simplify the 2D integration we transform the original integral into a contour integral using Green's theorem:

\begin{equation}
\iint_R \left( \frac{\partial Q}{\partial x} - \frac{\partial P}{\partial y} \right) \, dxdy = \oint_C \left( P \, dx + Q \, dy \right)
\end{equation}
where $R$ is the 2D integration region and $C$ is its edge. In Fig.~\ref{fig:cont_integ} we give an example of an integration contour for the edge of a lenslet. We need to find the functions $P(x,y)$ and $Q(x,y)$ that satisfy the following equation: $\frac{\partial Q}{\partial x} - \frac{\partial P}{\partial y} = f(x, y)$ where $f(x, y)$ is a model flux distribution. In our case, where $f(x,y)$ is a 2D Gaussian, we derive the following expressions for $P(x,y)$ and $Q(x,y)$ by solving first-order differential equation:

$$
P(x, y) = -\frac{1}{4 \sigma \sqrt{2 \pi}} \exp\left(-\frac{1}{2} \left( \frac{x - x_c}{\sigma} \right)^2 \right) 
\cdot \text{erf}\left(\frac{y - y_c}{\sigma\sqrt{2}} \right)
$$
$$
Q(x, y) = \frac{1}{4 \sigma \sqrt{2 \pi}} \exp\left(-\frac{1}{2} \left( \frac{y - y_c}{\sigma} \right)^2 \right) 
\cdot \text{erf}\left(\frac{x - x_c}{\sigma\sqrt{2}} \right)
$$
where $x_c$ and $y_c$ are the coordinates of the center of the Gaussian PSF. Using these expressions we have transformed the 2D integral to a 1D integral over a spaxel contour. This transformation allows us to reduce the dimensionality of the integration problem and therefore the overall computational time. 

To optimize the computation for the full lenslet array we can numerically integrate the vector functions $\vec{P(x,y)}$ and $\vec{Q(x,y)}$ where the number of elements corresponds to the number of spaxels. 

Analytic calculations for other PSF profiles commonly used in astronomy, e.g. the more realistic \citet{1969A&A.....3..455M} profile are harder, because the variables in the differential equations cannot be easily separated. Here, numerical 2D integration is required.

\end{appendix}

\end{document}